\documentclass[aps, prd, preprint, preprintnumbers, showpacs,showkeys, amsfonts, nofootinbib, floatfix]{revtex4-1}
\usepackage{graphicx}
\usepackage{epsfig}
\usepackage{rotate}
\usepackage{rotating}
\usepackage{multirow}
\usepackage{longtable}
\usepackage[T1]{fontenc}
\usepackage{array}

\begin{document}
\preprint{CUMQ/HEP 176}
\vspace{1.2cm}
\title{Sterile neutrinos in $U(1)'$ with $R$-parity Violation}

\author{Mariana Frank$^a$\footnote{mariana.frank@concordia.ca}}
\author{Levent Selbuz$^{a,b}$\footnote{levent.selbuz@eng.ankara.edu.tr}}
\affiliation{$^a$Department of Physics, Concordia University, 7141
Sherbrooke St. West, Montreal, Quebec, Canada H4B 1R6,}
\affiliation{$^b$Department of Engineering Physics, Ankara
University, TR06100 Ankara, Turkey}

\date{\today}

\begin{abstract}

Motivated by results from short-baseline neutrino oscillation data, we  study neutrino masses and mixing in $U(1)^\prime$ supersymmetric models with  $R$-parity breaking. Whether $R$-parity is broken spontaneous or through (effective) bilinear terms in the Lagrangian, the breaking terms  induce mixing between the neutralinos and neutrinos, creating a scenario in which some neutralinos can be heavy, and some light. Both the right-handed neutrino and the singlino (fermionic partner of the additional singlet Higgs field)  can be light, and act as sterile neutrinos,  which  reconcile some of the anomalies observed in solar baseline and reactor experiments. We show that, scanning a large range of the parameter space satisfying solar and atmospheric neutrino constraints, the mass and mixing parameters of the sterile neutrinos are very restrictive, leading to some predictive features for the $U(1)^\prime$ scenarios.

\end{abstract}

%%%%%%%%%%%%%%
%%%%%%%%%%%%%%
%\vspace*{-0.9cm}
\maketitle
%%%%%%%%%%%%%%

%%%%%%%%%%%%%%%%%%%%%%%%%%%%%%%%%%%%%%%%%%%%%%%%%%%%%%%%%%%%%%%

%\tableofcontents
\section{Introduction}
\label{sec:intro}

Solar, atmospheric, reactor and
accelerator neutrino experiments
have provided compelling evidence for neutrino
 ($\nu_e$, $\nu_{\mu}$, $\nu_{\tau}$) and antineutrino
 ($\bar{\nu_e}$, $\bar{\nu_{\mu}}$, $\bar{\nu_{\tau}}$) oscillations, all attributed to
nonzero neutrino masses and mixing.  The neutrino data,  accumulated over many
years,  has  allowed determination of the parameters in neutrino oscillations, $\Delta
m^2_{21}$, $\theta_{12}$, $|\Delta m^2_{31(32)}|$, and $\theta_{23}$, with a  high precision. Recent
 developments in March 2012 have resulted
 in a high precision determination of $\sin^2
2\theta_{13}$ by the Daya Bay experiment with reactor $\bar{\nu_e}$
\cite{An:2012eh}: $\sin^2 2\theta_{13}=0.089\pm0.010\pm0.005$. Subsequently, the
 Double Chooz \cite{Abe:2011fz},  T2K  \cite{Abe:2011sj} and RENO \cite{Ahn:2012nd}  experiments reported,
respectively, 4.9$\sigma$, 2.9$\sigma$ and 3.2$\sigma$ signals for
a non-zero value of $\theta_{13}$, compatible with the Daya Bay
result.

 A global analysis of the latest neutrino oscillation data
presented at the Neutrino 2012 International Conference, was
performed in \cite{Abazajian:2012ys}. The best fit values  obtained  are:
\begin{eqnarray}
\Delta m^2_{21}&=&7.54 \times 10^{-5}~{\rm eV}^2, \qquad
|\Delta m^2_{31(32)}|=2.47(2.46)\times 10^{-3}~{\rm eV}^2;\nonumber \\
\sin^2 \theta_{12}&=&0.307, \qquad \sin^2 \theta_{23}=0.39, \qquad \sin^2 \theta_{13}=0.0241(0.0244),
\end{eqnarray}
where the values (the values in brackets) correspond to $m_1 < m_2 <
m_3 (m_3 < m_1 < m_2)$, i.e., for the normal  (inverted) neutrino mass ordering.

The completeness of the three-neutrino mixing picture has been challenged by a re-evaluation of the expected electron antineutrino
($\bar{\nu_e}$) flux emitted at nuclear reactors \cite{Mention:2011rk}. The new
prediction is $\sim 3\%$ higher than previously assumed.
 If confirmed, this result would imply that all existing
neutrino oscillation searches at nuclear reactors have observed a
deficit of $\bar{\nu_e}$, which can be interpreted in terms of
oscillations at baselines of order $10$$-$$100$ m. But for
reactor antineutrino energies of a few MeV, standard oscillations of
the three active neutrinos require baselines of at least 1 km. Thus
the reactor anomaly can be accommodated only if (at least) one
sterile neutrino with mass at the eV scale or higher is introduced.
This is further supported by the long-standing LSND
anomaly \cite{Aguilar:2001ty},  the more recent MiniBooNE antineutrino
results \cite{AguilarArevalo:2010wv}, as well as by source calibrations performed for solar neutrino experiments based on gallium \cite{Hampel:1998xg}, all which  suggest the existence of a sterile neutrino in the $\Delta m^2_{41}>1~{\rm eV}^2$.

If a fourth type of neutrino has a mass not much larger
than the  three active neutrinos, the results of reactor neutrino oscillations
like  Daya Bay \cite{An:2012eh},  Double Chooz \cite{Abe:2011fz} and RENO \cite{Ahn:2012nd} can be affected by the fourth
state, as these detectors established oscillations driven by $\Delta
m^2_{31} = 0.00232$ eV$^2$.  The existence of a fourth neutrino can be
perceived only if the order of the mass splitting $\Delta m^2_{41}$ is not much larger
than that of $\Delta m^2_{31}$.

  In the last few years, a possible cosmological hint of
light sterile neutrinos (see e.g. \cite{Abazajian:2012ys} and references therein) was found by combining
the result in the best fit from WMAP, SDSS II-Baryon Acoustic
Oscillations, and Hubble Space Telescope data. The existence of light sterile neutrinos would have important consequences
for dark matter searches, Big Bang Nucleosynthesis, Cosmic Microwave Background, Hubble
constant and galaxy power spectrum. Although the significance of these hints for inert neutrinos depends on the data sample  and on assumptions inherent in cosmological models,  analyses favor the presence of light sterile neutrinos with mass-squared difference $\Delta m_{41}^2\sim 0.1 \,{\rm eV}^2$ \cite{Abazajian:2012ys}. Constraints on the number of neutrinos from Big Bang Nucleosynthesis
(BBN)  also allow for sterile neutrinos. Recent
analyses lead to $N_{eff}<4.26$ \cite{Hamann:2010pw} and $N_{eff}<4.1$ \cite{Mangano:2011ar} at 95\% C.L. limit, with best fit $N_{eff}=3.86$. As a result,  one fully thermalized eV sterile neutrino is preferred
by BBN, while two fully thermalized eV sterile neutrinos are disfavored.  The data from PLANCK satellite is consistent with the bound  $N_{eff}=3.30 \pm 0.27$ at 68\%C.L., with sterile-active mass-squared splittings in the range of $(10^{-5}-10^2)~{\rm eV}^2$ \cite{Planck}.

Inactive singlet neutrinos are familiar in the
seesaw mechanism \cite{Mohapatra:1979ia}, but they are extremely heavy.  But recent phenomenological studies  have been
performed in a framework in which the standard three active neutrino
scenario is amended by adding one $[3 + 1]$ or two $[3 + 2]$
sterile neutrinos with masses in the eV range. These studies suggest that an explanation of the  anomalies
within  sterile neutrino scenarios restricts severely the neutrino parameter space, and the excluded area covers the
region accessible by current and future laboratory experiments.  At present it seems that neither the scenario with, nor the one without, sterile neutrinos can explain {\it all} neutrino data \cite{Adamson:2011qu,Bodmann:1994py,Astier:2003gs,Antonello:2012pq}. However, the possibility of the existence of sterile neutrinos is theoretically interesting, as it would provide a signal for physics beyond the standard model (SM). Recent sterile neutrino reviews are available in  \cite{Giunti:2011bx}.

 We use this motivation to explore  breaking $R$-parity in supersymmetry (SUSY)  for generating sterile neutrino candidates. SUSY has the attractive feature that it provides the lightest supersymmetric particle (LSP) as the dark matter candidate. Breaking $R$-parity destroys this, as the LSP can now decay. However, the gravitino might still provide a dark matter candidate, as  it is sufficient to decay slowly enough to satisfy the relic density  \cite{Restrepo:2011rj}.
  We choose an extended supersymmetric scenario to highlight a new candidate for sterile neutrinos. The so-called $U(1)^\prime$ models, where the gauge symmetry of the SM is augmented by an extra $U(1)$ group, introduces an additional gauge boson and a (minimum) of one additional singlet scalar representation.   The $U(1)^\prime$ model provides a solution to  the $\mu$ problem in supersymmetry  \cite{muprob},  and forbids terms that violate baryon number from coexisting with lepton violating terms, making the proton stable. In models with broken $R$-parity, if lepton number is broken, the neutral (charged) Higgs mix with the sneutrinos (sleptons).  The neutralinos mix with the neutrinos (providing a mass mechanism for the neutrinos, whether Majorana or Dirac), and the charginos mix with the leptons. Higgs and collider phenomenology would be significantly affected \cite{inprogress}.  In this work, we concentrate on neutrino-neutralino mixing, and investigate the possibility that the sterile neutrino could be either one of the right-handed neutrinos, or the singlino (the fermionic partner of the singlet Higgs).  We forgo studies of $[3+2]$ scenarios, since the CP violating difference between neutrino and antineutrino in the MiniBooNE data is greatly decreased,  in favor of $[3+1]$.

Our work is organized as follows. First, we introduce our  model in Sec.  \ref{sec:model}, with emphasis on $R$-parity violation. We highlight  two sterile neutrino scenarios and the mixing between the light and heavy neutralinos. Within each scenario, we give approximate analytical expressions for mass eigenvalues of the neutrinos in the two scenarios.  We follow with the explicit numerical analysis and discussion of masses and mixings in Sec. \ref{sec:analysis}, then summarize our findings and conclude in Sec. \ref{sec:conclusion}.

%%%%%%%%%%%%%%%%%%%%%%%%%%%%%%%%%%%%%%%%%%%%%%%%%%%%%%%
\section{The model}
 \label{sec:model}
%%%%%%%%%%%%%%%%%%%%%%%%%%%%%%%%%%%%%%%%%%%%%%%%%%%%%%%
 In the $U(1)^\prime$ model \cite{Demir:2005ti}, the symmetry of the MSSM is augmented by an Abelian $U(1)$ group, and the particle spectrum is enlarged by an additional  neutral  gauge boson and at least one additional singlet Higgs representation, needed to break the extra symmetry.\footnote{To accommodate a large $Z-Z^\prime$ mass splitting, additional singlets are introduced into the theory. This variant of the model is known as the {\it secluded} $U(1)^\prime$ model \cite{Erler:2002pr}. For simplicity, we do not consider this scenario here, and restrict ourselves to one additional singlet.} The $U(1)^\prime$ charge assignments
which generate the $\mu_{eff}$ term induce mixed anomalies between the $U(1)^\prime$ and the $SU(3)_C \times SU(2)_L \times U(1)_Y$ groups. The cancellation of these anomalies requires introduction of  exotic fermions, vector-like with respect to the MSSM, but chiral under the $U(1)^\prime$ group. The price to pay for the introduction of these fields is loosing the successful gauge coupling unification, which is an achievement of supersymmetry.

The superpotential of the model is given by
\begin{eqnarray}\label{eq:superpot}
\widehat{W}_{U(1)^\prime}&=&{\mathbf h}_u\widehat{Q}\cdot \widehat{H}_u \widehat{U}+
{\mathbf h}_d\widehat{Q}\cdot \widehat{H}_d \widehat{D} + {\mathbf h}_e\widehat{L}\cdot
\widehat{H}_d \widehat{E} + {\mathbf h}_s \widehat{S}\widehat{H}_u \cdot
\widehat{H}_d +    \widehat{L}\cdot
\widehat{H}_u {\mathbf h}_{\nu} \widehat{N}
\\
&+& \sum_{i=1}^{n_{\cal{Q}}} {h}_Q^i \widehat{S} \widehat{\cal{Q}}_i
\widehat{\cal{\overline{Q}}}_i + \sum_{j=1}^{n_{\cal{L}}} {h}_L^j
\widehat{S} \widehat{\cal{L}}_j \widehat{\cal{\overline{L}}}_j,
\end{eqnarray}
  where $\widehat{\cal {L}}$ and $\widehat{\cal {Q}}$ are the exotic messenger fields, and
${\mathbf h}_{\nu}$ is the Yukawa coupling responsible for generating neutrino
masses.

 In the above equation, ${\mathbf h}_\nu\sim \lambda/M_R$, with $\lambda$ a coupling of the same order of the Yukawa couplings, and $M_R$ a large mass scale.  The couplings  ${\bf h_u}, {\bf h_d}, {\bf h_e}$ represent the usual quark and lepton Yukawa matrix couplings, while $h_s$ is the singlet coupling with the MSSM Higgs doublets. The Abelian gauge symmetry $U(1)^\prime$ is assumed to be broken at higher scales by the vacuum expectation value (VEV) of the singlet Higgs field $\langle S \rangle=\frac{v_S}{\sqrt{2}}$. This VEV also yields an effective $\mu$ term dynamically, $\mu_{eff}=\displaystyle h_s{\langle S \rangle}$. The $SU(2)_L \times U(1)_Y$ is broken as usual by the VEVs of the Higgs doublets
\begin{equation}
\langle H_u \rangle=\frac{1}{\sqrt{2}} \left ( \begin{array}{c} 0 \\ v_u \end{array} \right ),  \qquad \langle H_d \rangle=\frac{1}{\sqrt{2}} \left ( \begin{array}{c} v_d \\ 0 \end{array} \right ).
\end{equation}

%The aim of this study is to keep the model as general as possible. 
In order to solve the $\mu$-problem and to allow for a singlet field whose fermionic partner can serve as a sterile neutrino, an additional $U(1)^\prime$ gauge group is needed. The model can be NMSSM, but we have chosen the $U(1)^\prime$ model for its attractive features. 
%Further details of the model, including the pattern of symmetry breaking and a complete set of anomaly cancellation conditions are provided in \cite{Demir:2010is}.

If $R$-parity breaking is allowed, this would induce additional terms in the Lagrangian.  In $U(1)^\prime$, $R$-parity can be broken in two different ways:
\begin {enumerate}
\item The Lagrangian can contain explicit  $R$-parity breaking terms as in \cite{Lee:2007fw},  for lepton  number violation:
\begin{equation}
\widehat{W}_{LV}=\varepsilon _{ab}h_{s,i}^\prime S H^a_uL^b_i+\lambda_{ijk}L_iL_jE^c_k +\lambda^{\prime}_{ijk}L_iQ_jD^c_k,
\label{bilinear}
\end{equation}
or for baryon number violation:
\begin{equation}
\widehat{W}_{BV}=\lambda^{\prime \prime}_{ijk}U^c_iD^c_jD_k^c.
\end{equation}
The first term in Eq. (\ref{bilinear}) is the so-called bilinear term $\mu_i^{\prime \prime} H_u L_i$, but here $\mu_{i\,eff}^{\prime  \prime}$ is promoted to a dynamical variable when
$U(1)^\prime$ is broken, $\mu_{i, eff}^{\prime \prime
}=h^{\prime}_{s,i}\langle S \rangle$. As was shown before \cite{Lee:2007fw}, in $U(1)^\prime$, unlike in the minimal model, one can have {\it either} explicit lepton {\it or} explicit baryon number violating interactions, {\it but not both}, thus forbidding proton decay which requires both terms be non-zero. Additionally $U(1)^\prime$ symmetry disallows higher-dimensional proton decay inducing operators which are otherwise expected to appear at a higher scale. Thus, in $U(1)^\prime$ models with $R$-parity violation, the $\mu$ problem is solved, and the proton is stable, even in the presence of exotic fields required for anomaly cancellation  \cite{Lee:2007fw}.
\item $R$-parity violation can occur spontaneously, through sneutrinos acquiring a VEV: $\langle \widetilde{\nu}_{\{L,R\}_i}\rangle = v_{\{L,R\}_i}$, as in  \cite{Hirsch:2008ur}.
This breaking has a similar effect as allowing bilinear $R$-parity violating terms, but it induces additional mixing
between neutral gauginos and neutrinos, and these are important in our considerations, as we shall see later. Assuming that the
 sneutrino VEV is $\langle \widetilde \nu_{L_i }\rangle \sim 10^{-4}$
GeV would not interfere with small neutrino masses \cite{Demir:2007dt}.
\end{enumerate}

As we wish to investigate the effect of the $R$-parity violation in the neutralino
sector of the $U(1)^\prime$, we restrict ourselves to the case of spontaneous and bilinear $R$-parity violation, that is $\widehat{W}_{\not{R}}=\widehat{W}_{U(1)^\prime}+\varepsilon _{ab}h_{s,i}^\prime S H^a_uL^b_i$. We keep the particle content of the model as minimal as possible, while insuring the existence of the  $\mu_{eff}$ term, proton stability and neutrino masses.
The complete Lagrangian of this model 
incorporates kinetic terms and various interaction terms among the fields. 
The kinetic terms of the Lagrangian are given by
\begin{eqnarray}\label{kinetic}
{\cal L}_{U(1)^\prime}^{Kinetic }&=&{\cal
L}_{MSSM}^{Kinetic}-\frac{1}{4}Z^{\prime\mu\nu}Z^\prime_{\mu\nu}+({\cal
D}_{\mu}S)^{\dagger}({\cal D}^{\mu}S)+\tilde Z^{\prime
\dagger}i\sigma^{\mu}\partial_{\mu}\tilde Z^{\prime}+\tilde
S^{\dagger}i\sigma^{\mu}{\cal D}_{\mu}\tilde S+({\cal
D}_{\mu}\tilde N)^{\dagger}({\cal D}^{\mu}\tilde N), \nonumber\\
\end{eqnarray}
where $j=1,2,3$. The interactions of the gauge fields with the rest
(fermions,
sfermions, gauginos, Higgs and Higgsino fields) are contained in the
piece
\begin{eqnarray}
{\cal{L}}^{gauge}_{U(1)^{\prime}} = {\cal{L}}^{gauge}_{MSSM}\left(g_Y
\frac{Y_{X}}{2} B_{\mu} \rightarrow g_Y \frac{Y_{X}}{2}
B_{\mu} + g_{Y^{\prime}} Q_X^{\prime} Z^{\prime}_{\mu} \right)\,,
\end{eqnarray}
where $Y$ is the hypercharge and $X$ runs over the fields charged under $U(1)^{\prime}$. In
(\ref{kinetic}), $Z^{\prime\, \mu \nu}$ is the field strength tensor
of $Z^{\prime}_{\mu}$, and ${\cal
D}_{\mu}S_j=(\partial_{\mu}+ig_{Y^{\prime}}Q^{\prime}_{S_{j}}Z_{\mu}^{\prime}
)S_j$ for $j=1,2,3$.

The soft-breaking sector of the $U(1)^{\prime}$ Lagrangian is
\begin{eqnarray}\label{soft}
{\cal L}_{U(1)^\prime}^{Soft}&=&{\cal
L}_{MSSM}^{Soft
}(\mu\rightarrow0)-m^2_SS^*S -m_{\tilde L}^2  |\tilde L|^2-m^2_{\tilde E^c}|\tilde E^c|^2
-m^2_N\tilde N^*\tilde N +\frac{1}{2}\biggl(M_{\tilde Z^\prime}\tilde Z^\prime\tilde
Z^\prime+h.c.\biggr)\nonumber \\
&-&[h_sA_sSH_u\cdot H_d+{\mathbf h}_{\nu}A_{\nu} \tilde L\cdot
H_u\tilde N+h.c.]\, ,
\end{eqnarray}
where $M_{\widetilde{Z}^{\prime}}$ is $U(1)^{\prime}$ gaugino mass, and $A_s$
is the extra trilinear soft coupling.

The  $F$--terms in the $U(1)^{\prime}$ Lagrangian are given
by
\begin{eqnarray}\label{Fterm}
{\cal L}_{U(1)^\prime}^{F-term }&=&-\sum_i\left|\frac{\partial
W}{\partial\phi_i}\right|^2={\cal L}_{MSSM}^{F-term }(\mu\rightarrow
h_sS)-h_s^2|H_u\cdot H_d|^2-\left({\mathbf h}_u\tilde Q^*\tilde
U^*+h_s^*S^*H_d^*\right){\mathbf h}_{\nu}\tilde L\tilde N\nonumber \\
&-&{\mathbf h}_{\nu}\tilde L^*\tilde N^*\left(h_u\tilde Q\tilde
U+h_sSH_d+{\mathbf h}_{\nu}\tilde L\tilde N\right)-
\left({\mathbf h}_eH_d^*\tilde E^*\right)\left ({\mathbf h}_{\nu}H_u\tilde N \right) \nonumber \\
&-&{\mathbf h}_{\nu}H_u^*\tilde N^*\left({\mathbf h}_eH_d\tilde
E+{\mathbf h}_{\nu}H_u\tilde N\right)-
{\mathbf h}_{\nu}^2|\tilde L\cdot H_u|^2 \, ,
\end{eqnarray}
where $\phi_i$ is the scalar component of the $i$--the chiral superfield in the
superpotential.

The $D$--term contributions to the Lagrangian are given by
\begin{eqnarray}\label{Dterm}
{\cal L}_{U(1)^\prime}^{D-term }&=&-\frac{1}{2}\sum_a D^a D^a=-\frac{(g_Y^2+g_2^2)}{2}\left (|H_u^0|^2 - |H_d^0|^2- \sum_i|\tilde \nu_i|^2 \right)^2-
\frac{g_{Y^\prime}^2}{2}\biggl(Q^{\prime}_Q\tilde Q^{*}\tilde
Q  +Q^{\prime}_U\tilde U^{*}\tilde U\nonumber \\
&+&Q^{\prime}_D\tilde D^{*}\tilde
D
+Q^{\prime}_L\tilde L^{*}\tilde L+Q^{\prime}_E\tilde E^{*}\tilde
E+Q^{\prime}_{H_d}H_d^{*}H_d+Q^{\prime}_{H_u}H_u^{*}
H_u+Q^{\prime}_N\tilde
N^*\tilde N+Q^{\prime}_SS^*S \biggr)^2\, .
\end{eqnarray}
Through electroweak symmetry breaking one gauge boson remains massless ($\gamma$) and two become massive ($Z, \,Z^\prime$).  Thus, in addition to the new singlet Higgs field, the model introduces a new neutral gauge boson, $Z^\prime$ at the $U(1)^\prime$ symmetry breaking scale, with mass \cite{Demir:2005ti,Lee:2007fw}
\begin{equation}
M^2_{Z^\prime}=g^2_{Y^\prime} \left [ Q^{\prime\,2}_{H_d} v_d^2+ Q^{\prime\,2}_{H_u} v_u^2+ Q^{\prime\,2}_{S} v_S^2 +\sum_{i=1}^3Q^{\prime \,2}_{N_i}v_{R_i}^2\right ], 
\end{equation}
where $Q^\prime_i$ are the charges of the particle $i$ under the $U(1)^\prime$ group,  and $g_{Y^\prime}$  is the $U(1)^\prime$ gauge coupling constant. The mass of   the $Z^\prime$ boson depends on the  VEV of the singlet field, and the $U(1)^\prime$ charges for the Higgs fields and right-handed sneutrinos. As this boson is not seen at the LHC, it is an indication that i) either the scale of the model, $\langle S \rangle$, is high, ii) some of the right-handed sneutrino VEVs are large, or iii) a secluded sector is needed to generate a large $Z-Z^\prime$ mass splitting.  We adopt the second assumption, that is, for this model $v_S >1$ TeV,  and could, in practice, be allowed to be much larger; and $v_{R_i} \gg 1$ TeV for some $i$. Exotic matter multiplets are introduced into the Lagrangian for anomaly cancellation, which depends entirely on their $SU(3)_c \times SU(2)_L \times U(1)_Y \times U(1)^\prime$ charge assignments and not on their mass. They must have the same $U(1)^\prime$ charge as $H_u$ and $H_d$ to allow for the coupling to the $S$ scalar. Thus their mass is generated by the VEV $v_S$, and in principle, they would be expected to have masses of that order. We work in the large $v_S$ scenario, $v_S \gg v_u, v_d$. We shall see in Section \ref{sec:analysis} that there are two scenarios possible: one in which $v_S \sim 1$ TeV, and the  exotics have mass in the TeV region; and another when $v_S\sim 10^6$ TeV, and the exotics will be similarly heavy.  In addition, there is some choice in selecting the $U(1)^\prime$ charge assignments of the exotic fields, so that the lowest dimension operators coupling the exotics to the MSSM fields are absent, making the former difficult to observe \cite{Lee:2007fw}. A complete set of anomaly cancellation conditions exist in the literature \cite{Demir:2005ti} and we do not repeat them here. However,  for consistency we list the particle content and charges of the $U(1)^\prime$ model in Table \ref{tab:charge-sol}.

\begin{table}[t]
\begin{tabular*}{0.99\textwidth}{@{\extracolsep{\fill}} cccc}
\hline\hline $\displaystyle
\begin{array}{l} Q^\prime_{H_u}=-2\\
Q^\prime_{H_d}=1\\
Q^\prime_{S}=1\\
Q^\prime_{Q}=x\\
Q^\prime_{U}=2-x
\end{array}$&$\begin{array}{l} Q^\prime_{D}=-1-x\\
Q^\prime_{L}=\frac{1}{3}-3x\\
Q^\prime_{E}=-\frac{4}{3}+3x\\
Q^\prime_{N}=\frac{5}{3}+3x
\end{array}$&$\begin{array}{l} Q^\prime_{\cal{Q}}=
\frac{4-12x-\sqrt{2}\Omega}{18}\\
Q^\prime_{\cal{\overline{Q}}}=\frac{-22+12x+\sqrt{2}\Omega}{18}\\
Q^\prime_{\cal{L}}=\frac{-15+13\sqrt{10}-12\sqrt{10}x+\sqrt{5}\Omega}{30}\\
Q^\prime_{\cal{\overline{L}}}=\frac{-15-13\sqrt{10}+
12\sqrt{10}x-\sqrt{5}\Omega}{30}
\end{array}$\\ \hline\hline
\end{tabular*}
\caption{\label{tab:charge-sol}\sl\small A set of $U(1)^{\prime}$ charges
satisfying all gauge invariance and anomaly cancellation
conditions. The charge of the quark and lepton doublets  depend on the parameter $x$ and for the exotics $\widehat{\cal {Q}}$ and $\widehat{\cal {L}}$ the parameter 
 $\Omega(x)=\sqrt{241+708x+612x^2}$ is
introduced.}
\end{table}

 With $R$-parity conservation, the
$U(1)^{\prime}$ model  has two additional (with respect to MSSM) fermion fields in the
neutral sector: the $U(1)^{\prime}$ gauge fermion
$\widetilde{Z}^{\prime}$ and one singlino $\widetilde{S}$,
 in total,
six neutralino states $\widetilde{\chi}_i^0$ ($i=1,\dots,6$). If the 
$R$-parity is broken, the neutralino and neutrino states mix, and we
have additionally, three left-handed, and three right-handed states in the neutralino mass matrix. In what follows,  we restrict ourselves to one right-handed neutrino (chosen to be the right-handed $\tau$ neutrino, $N_\tau$), for simplicity. We assume the others to be very heavy and decouple from the rest of the neutralino spectrum.

We work in the basis: $\widetilde {\psi}_i= (\nu_e,\, \nu_\mu,\, \nu_\tau, N_\tau, \widetilde{B},  \widetilde
W, \widetilde {B}^\prime, \widetilde {H}_d, \widetilde {H}_u, \widetilde{S}$), where the neutralino mass matrix is
\begin{eqnarray}
\label{N.Mass}
\mathcal{M} =\left(
 \begin{array}{cccccccc}
    0_{3 \times 3}
    &
    h_{\nu_j}v_{u}
    &
    -\frac{g_{Y} v_{Lj}}{2}
    &
    -\frac{g_2 v_{Lj}}{2}
    &
    - \frac{g_{Y^\prime} v_{Lj}}{2}
    &
    0_{3 \times 1}
    &
    h_{\nu_j} v_R+\mu_j^{\prime \prime}
    &
    0_{3 \times 1}
\\
    h_{\nu}v_{u}
    &
    0
    &
    0
    &
    0
    &
    \frac{g^\prime_{1} v_{R}}{2}
    &
    0
    &
    h_\nu v_{Lj}
    &
    0
\\
    -\frac{g_{Y}  v_{Li}}{2}
    &
    0
    &
    M_{\tilde Y}
    &
    0
    &
    M_{\tilde Y \tilde Y^\prime}
    &
    -\frac{g_Yv_d}{2}
    &
    \frac{g_Yv_u}{2}
    &
    0
\\
    -\frac{g_{2} v_{Li}}{2}
    &
    0
    &
    0
    &
    M_{\tilde W}
    &
    0
    &
    \frac{g_2 v_d}{2}
    &
    -\frac{g_2 v_u}{2}
    &
    0
\\
    -\frac{g_{Y^\prime} v_{Li}}{2}
    &
    \frac{g^\prime_{1} v_{R}}{2}
    &
    M_{\tilde Y \tilde Y^\prime}
    &
    0
    &
    M_{\tilde Y^\prime}
    &
    \mu^\prime_{d}
    &
    \mu^\prime_{u}
    &
    \mu^\prime_S
\\
    0
    &
    0
    &
    -\frac{g_Y v_d}{2}
    &
    \frac{g_2 v_d}{2}
    &
    \mu^\prime_{d}
    &
    0
    &
    -\mu
    &
    -\mu_{d}
\\
    h_{\nu_i} v_R+\mu_i^{\prime \prime}
    &
    h_\nu v_{Li}
    &
    \frac{g_Y v_u}{2}
    &
    -\frac{g_2 v_u}{2}
    &
    \mu^\prime_{u}
    &
    -\mu
    &
    0
    &
    -\mu_{u}
\\
    0_{1 \times 3}
    &
    0
    &
    0
    &
    0
    &
    \mu_S^\prime
    &
    -\mu_{d}
    &
    -\mu_{u}
    &
    0
 \end{array}
 \right).
 \label{eq:allneutralinos}
 \end{eqnarray}
where we used the notation:  $\displaystyle \mu_{d}=h_s\frac{v_d}{\sqrt{2}}, \, \mu_{u}=h_s\frac{v_u}{\sqrt{2}}, \, \mu=h_s \frac{v_S}{\sqrt{2}}, \, \mu^\prime_{d}=g_{Y^\prime} Q^\prime_{H_d}v_d, \,\mu^\prime_{u}=g_{Y^\prime} Q^\prime_{H_u}v_u$ and $\mu^\prime_{S}=g_{Y^\prime} Q^\prime_{S}v_S, \, \mu_j^{\prime \prime}=h^\prime_{s, j} \frac{v_S}{\sqrt{2}},$, with $Q^\prime_i$ the corresponding charges under the $U(1)^\prime$ group. We also denoted $g_1^\prime=g_{Y^\prime}\frac{Q_{N}^\prime}{Q_L^\prime}$. In the above matrix, $M_{\tilde Y},~M_{\tilde W}$ and $M_{\tilde Y^\prime}$ are the $U(1),~SU(2)_L$ and $U(1)^\prime$ gaugino masses, and $M_{\tilde Y \tilde Y^\prime}$ is the $U(1)-U(1)^\prime$ mixing mass parameter.

In what follows, we shall consider scenarios where the three active neutrinos and one additional one (chosen to serve as sterile neutrino) are light, while the rest of neutralinos are heavy, effectively acting as a seesaw mechanism. We explore two candidates for sterile neutrinos.  In one scenario, the right-handed $\tau$ neutrino is the sterile candidate. In the other scenario, the right-handed neutrinos are all heavy, while the singlino is  light and acts as  sterile neutrino.

\subsection{The first [3+1] Scenario: the right-handed tau neutrino as sterile neutrino}

 Using right-handed neutrinos as light sterile neutrinos was considered before. In particular, this is a consequence of breaking of $U(1)_{B-L}$ \cite{Barger:2010iv,Ghosh:2010hy,Duerr:2013opa}, but light right-handed sterile neutrinos also appear in the context of split seesaw models \cite{Kusenko:2010ik}, and in  string theories \cite{Braun:2005nv}.  

In this scenario we take
only $N_\tau$ to be the {\it light} right-handed neutrino, and add it as well as the left-handed neutrinos
$\nu_e,\, \nu_\mu,\, \nu_\tau$, to the neutralino mass matrix.  For this, we need to invoke a hierarchy in neutrino masses, so that one right-handed neutrino is light, while the other two remain heavy. The possibility of an existing symmetry within  type I seesaw models  which can accommodate  one  light right-handed neutrino  was pointed out in  \cite{Petcov:1982ya}. A popular method is based on  the $L_e-L_\mu-L_\tau$
symmetry, which  leads to a very characteristic mass pattern for active
neutrinos, in which  one neutrino is exactly massless. Applying the same 
symmetry to three right-handed neutrinos yields an analogous pattern
$(0,M,M)$ for the heavy neutrino masses. This symmetry must be further broken  very weakly by loop or by higher dimensional operators, lifting the degeneracies and giving mass to the massless particle, chosen to be the light sterile neutrino \cite{Lindner:2010wr}.  The symmetry breaking scale $\Lambda \sim \langle v_{R} \rangle $ must be smaller than the preserving scale $M_{U(1)'} \sim \langle S \rangle$.  This mechanism has been used to motivate a small scale for the VEV of the right-handed sneutrino \cite{Barger:2010iv}. While the procedure was applied for $U(1)_{B-L}$,  the general features hold for our model as well. As the method is essential for our scenario, we outline it here.

Consideration of the terms in the potential  indicate that the dominant part comes from $D$-terms and soft-terms associated with the right-handed sneutrino neglecting terms in $h_\nu$ and $\langle \tilde \nu \rangle$ as much smaller), 
\begin{equation}
V \subset \,+\frac{g_{Y^\prime}^2}{2}\left (Q^{\prime}_N\tilde
N^*\tilde N +V_{MSSM}^{D-term}\right )^2+ m^2_N\tilde
N^*\tilde N
\end{equation}
Minimizing with respect to the sneutrino VEVs, a possible solution is the one in which only one right-handed sneutrino VEV is non-zero. This leaves one right-handed neutrino to get a TeV scale mass, leaving the other masses to be determined by the model parameters responsible for generating light active neutrino masses. The method requires a tachionic mass term for the right-handed sneutrino $m_N^2<0$, but allows at least one very light right-handed sneutrino VEV. 

In what follows, we use the procedure of \cite{Barger:2010iv}, but allow two right-handed neutrinos to be heavy. In order to use a seesaw mechanism, the VEV of the right-handed sneutrino must be light, and we expect the VEV of the singlet $v_S$ to be in the  TeV range,  the scale at which $U(1)^\prime$ symmetry will break.

%We defined $\nu_i=\langle \widetilde{\nu}_i \rangle$ and $M_N$ is the Majorana mass of the right-handed $\tau$ neutrino and $m_D^i=\frac{1}{\sqrt{2}}h_\nu^iv_u$. Note that trilinear terms do not give any mixing. In that case, the neutrinos get masses through loop diagrams with leptons/charginos  or quarks/charginos in the loop.

 The mass matrix $\mathcal{M}$ in Eq. (\ref{eq:allneutralinos}) contains six heavy states which can be integrated out using the seesaw mechanism to yield light neutrino masses (three active and one sterile)
$$\mathcal{M}_\nu=m_\nu-m_DM^{-1}m_D^T,$$
where the light $4 \times 4$ Majorana neutrino mass is given by:
\begin{eqnarray}
\label{nu.Mass}
m_\nu =\left(
 \begin{array}{cc}
    0_{3 \times 3}& h_{\nu_i}v_{u}\\
    h_{\nu_i}v_{u} & 0
    \end{array}
    \right), 
    \end{eqnarray}
the Dirac mass matrix is:
\begin{eqnarray}
\label{D.Mass}
m_D =\left(
 \begin{array}{cccccc}
    -\frac{g_{Y} v_{Li}}{2}
    &
    -\frac{g_2 v_{Li}}{2}
    &
    -\frac{g_{Y^\prime} v_{Li}}{2}
    &
    0_{3 \times 1}
    &
    h_{\nu _i} v_R+ \mu_i^{\prime \prime}
    &
    0_{3 \times 1}\\
     0
%    &0
     &
    0
    &
    \frac{g^\prime_{1} v_{R}}{2}
    &
    0
    &
    h_{\nu_\tau} v_{L\tau}
    &
    0
    \end{array}
    \right), 
    \end{eqnarray}
and the heavy Majorana neutralino mass matrix is, in the ( $\widetilde{B},  \widetilde
W, \widetilde {B}^\prime, \widetilde {H}_d, \widetilde {H}_u, \widetilde{S}$) basis:
\begin{widetext}
\begin{eqnarray}
\label{Maj.Mass}
M =\left(
 \begin{array}{cccccc}
    M_{\tilde Y}
    &
    0
    &
    M_{\tilde Y \tilde Y^\prime}
    &
    -\frac{g_Yv_d}{2}
    &
    \frac{g_Yv_u}{2}
    &
    0
\\
    0
    &
    M_{\tilde W}
    &
    0
    &
    \frac{g_2 v_d}{2}
    &
    -\frac{g_2 v_u}{2}
    &
    0
\\
    M_{\tilde Y \tilde Y^\prime}
    &
    0
    &
    M_{\tilde Y^\prime}
    &
    \mu^\prime_{d}
    &
    \mu^\prime_{u}
    &
    \mu^\prime_S
\\

    -\frac{g_Y v_d}{2}
    &
    \frac{g_2 v_d}{2}
    &
    \mu^\prime_{d}
    &
    0
    &
    -\mu
    &
    -\mu_{d}
\\

    \frac{g_Y v_u}{2}
    &
    -\frac{g_2 v_u}{2}
    &
    \mu^\prime_{u}
    &
    -\mu
    &
    0
    &
    -\mu_{u}
\\

    0
    &
    0
    &
    \mu_S^\prime
    &
    -\mu_{d}
    &
    -\mu_{u}
    &
    0

 \end{array}
 \right).
\end{eqnarray}
\end{widetext}
Inverting the neutralino mass matrix $M$ cannot be performed analytically in a closed form, and even when this is possible, such as for the case of a $4 \times 4$ matrix, the results are cumbersome and thus not particularly illuminating. For the purpose of this calculation, approximate analytical results are obtainable for the case of weak coupling, that is, assuming the couplings of the MSSM higgsino doublets to the singlet higgsino and $U(1)^\prime$ gaugino, as well as the couplings of the $U(1)_Y$ and $U(1)^\prime$ gaugino singlets, to be weak. One could then obtain an approximate solution following the procedure of \cite{Choi:2006fz}. We assume here that the neutralino mass matrix is real, and all the gaugino mass parameters are much larger than the electroweak and mixing scales, that is $M_{\tilde Y}, M_{\tilde W}, M_{\tilde Y^\prime} \gg \mu, \mu^\prime, M_{\tilde Y \tilde Y^\prime}$.

The neutrino mass matrix then becomes:
\begin{equation}
\label{Mnu}
\mathcal{M}_\nu=\left( \begin{array}{cc}
    A_1 \left [v_{Li} v_{Lj}\right ]/4
    + B_1 \left[(h_\nu^{\prime \prime} )_{i}v_{Lj} \right]/2
    + C_1\left[ (h_\nu^{\prime \prime} )_{i} (h_\nu^{\prime \prime} )_{j}\right]
    &
    \left [v_u h_\nu \right] _{j }/2
\\
    \left [v_u h_\nu\right]_{ i} /2
    &
    D_1 \left[  g_1^\prime v_{R} \right]^2 /4
\end{array}
\right)
\end{equation}
where we used the notation $h_\nu^{\prime \prime }\equiv h_\nu v_{R}+\mu^{\prime \prime}$, and where, to first order in small mixing parameters, we have
\begin{eqnarray}
    A_1 & =& \frac{g_Y^2}{M_{\tilde Y}}\left (1-\frac{M_Z^2 s_W^2}{M_{\tilde Y}^2}\right) +  \frac{g_2^2}{M_{\tilde W}}\left (1-\frac{M_Z^2 c_W^2}{M_{\tilde W}^2}\right)+
     \frac{g_{Y^\prime}^2}{M_{\tilde Y^\prime}}\left (1-\frac{Q^{\prime \, 2}_S m_s^2}{M_{{\tilde Y}^\prime}^2}\right) , \\
    B_1 & =& -\frac{2g_Y}{M_{\tilde Y}}  \frac{Q_+^\prime m_v M_{\tilde Y \tilde Y^\prime} }{M_{\tilde Y^\prime}(M_{\tilde Y}-M_{\tilde Y^\prime})}\left [-\frac{M_Y}{\mu}+M_Z^2\left( \frac{s_W^2}{M_{\tilde Y}^2}+ \frac{M_{\tilde Y} c_W^2+M_{\tilde W}s_W^2}{2\mu^2 M_{\tilde W}}(1+\sin 2 \beta)\right) \right ]  \nonumber \\
    & &- \frac{2g_{Y^\prime}}{M_{\tilde Y^\prime}} \frac{Q_2^\prime m_v \sin \beta}{\mu } \left [ 1+  \frac{M_Z^2(M_{\tilde Y} c_W^2+M_{\tilde W}s_W^2)}{\mu M_Y M_{\tilde W}} \right],  \\
    C_1 &=& \frac{1}{\mu}+  \frac{M_Z^2(M_{\tilde Y} c_W^2+M_{\tilde W}s_W^2) }{\mu^2 M_Y M_{\tilde W}}, \\
    D_1  &=& \frac{1}{ M_{\tilde Y^\prime} }- \frac { M_{\tilde Y \tilde Y^\prime}^2} {(M_{\tilde Y}-M_{\tilde Y^\prime})^2} \left ( \frac{1}{M_{\tilde Y^\prime}} +\frac {M_Z^2 s_W^2}{M_{\tilde Y}^3} -\frac{1}{M_{\tilde Y}} \right)   ,
\end{eqnarray}
with
\begin{eqnarray}
Q_+^\prime m_v&=&\frac{g_{Y^\prime}v}{\sqrt{2}} \left[\frac{Q^\prime_{H_d}\cos \beta+Q_{H_u}^\prime \sin \beta}{\cos\chi} +\frac{g_Y \tan\chi (\cos \beta-\sin \beta)}{2 g_{Y^\prime}}\right ], \nonumber\\
Q_2^\prime &=& \frac{Q^\prime_{H_u} }{\cos\chi} -\frac{g_Y \tan\chi}{2 g_{Y^\prime}}.
\end{eqnarray}
Here $\chi $ is the kinetic mixing angle, normally assumed to be $\sin\chi= 5 \times 10^{-3}$ \cite{Ali:2009md}. We have defined in the above, $m_s=g_{Y^\prime} v_S,~m_v=g_{Y^\prime} v$ and $s(c)_W=\sin(\cos) \theta_W$.

%%%%%%%%%%%%%%%%%%%%%%%%%%%%%%%%%%%%%%%%%%%%%%%%%%%%%%%%%%%%%%%%%%%%%%%%%%%%%%

\subsection{The second [3+1] Scenario: the singlino as sterile neutrino}
%%%%%%%%%%%%%%%%%%%%%%%%%%%%%%%%%%%%%%%%%%%%%%%%%%%%%%%%%%%%%%%%%%%%%%%%%%%%%%%%%%%%%%%%
 In the second scenario, we designate  the singlino, assumed to be light, to be the sterile neutrino. In this scenario, we can expect the VEVs of the right-handed sneutrinos to be heavy. We do not impose a scale for the VEV of the singlet field $S$, responsible for breaking the $U(1)^\prime$ symmetry, but rather fit it to yield the singlino as a light sterile neutrino.
 
 The singlino is in a unique position to act like a sterile neutrino. It is, as required, a neutral particle with no ordinary weak interactions except those induced by mixing, and although is it not a lepton, it acquires neutral lepton-like properties from mixing with the neutrinos. By comparison, the $U(1)^\prime$ bino mixes with the other gauginos and interacts via the gauge sector with quarks and leptons.  Thus a  model which accommodates the singlet naturally would contain another $U(1)$ gauge group beyond MSSM.  The NMSSM satisfies the requirement, but we choose the $U(1)^\prime$ model as it evades the domain-wall problem associated with the NMSSM \cite{Abel:1995wk}.
 
 In this case the mass matrix $\mathcal{M}$ in Eq. (\ref{eq:allneutralinos}) again contains six heavy states which can be integrated out using the seesaw mechanism to yield light neutrino masses (three active and one sterile)
$$\mathcal{M}_\nu=-m_DM^{-1}m_D^T,$$
as the light Majorana neutrino mass is $
m_\nu =\left(
    0_{4 \times 4}
    \right)$.
The Dirac mass matrix is:
\begin{eqnarray}
\label{D2.Mass}
m_D =\left(
 \begin{array}{cccccc}
 h_{\nu_j}v_u&
    -\frac{g_{1} v_{Li}}{2}
    &
    -\frac{g_2 v_{Li}}{2}
    &
    -\frac{g_{Y^\prime} v_{Li}}{2}
    &
    0_{3 \times 1}
    &
    h_{\nu} v_R+ \mu_i^{\prime \prime}
    \\
    0
    &
    0
    &
    0
    &
    \mu_S^\prime
    &
    -\mu_{d}
    &
    -\mu_{u}
    \end{array}
    \right), 
    \end{eqnarray}
and the heavy Majorana neutralino mass matrix is, in the ($N_\tau, \widetilde{B},  \widetilde
W, \widetilde {B}^\prime, \widetilde {H}_d, \widetilde {H}_u$) basis:
\begin{widetext}
\begin{eqnarray}
\label{Maj.Mass2}
M =\left(
 \begin{array}{cccccc}
    0
    &
    0
    &
    0
    &
    \frac{g^\prime_{1} v_{R}}{2}
    &
    0
    &
    h_\nu v_{Lj}
    \\
    0&  M_{\tilde Y}
    &
    0
    &
    M_{\tilde Y \tilde Y^\prime}
    &
    -\frac{g_Yv_d}{2}
    &
    \frac{g_Yv_u}{2}

\\
    0
    &
    0
    &
    M_{\tilde W}
    &
    0
    &
    \frac{g_2 v_d}{2}
    &
    -\frac{g_2 v_u}{2}

\\
\frac{g_1^\prime v_{R}}{2}
&
    M_{\tilde Y \tilde Y^\prime}
    &
    0
    &
    M_{\tilde Y^\prime}
    &
    \mu^\prime_{d}
    &
    \mu^\prime_{u}

\\
    0
    &
    -\frac{g_Y v_d}{2}
    &
    \frac{g_2 v_d}{2}
    &
    \mu^\prime_{d}
    &
    0
    &
    -\mu

\\
    h_{\nu}v_{Li}
    &
    \frac{g_Y v_u}{2}
    &
    -\frac{g_2 v_u}{2}
    &
    \mu^\prime_{u}
    &
    -\mu
    &
    0
 \end{array}
 \right).
\end{eqnarray}
\end{widetext}
Notice that in this case, consideration of spontaneous $R$-parity violation is important for right-handed neutrino-higgsino mixing, as otherwise the right-handed neutrino would decouple from the spectrum.
The neutrino mass matrix then becomes, as an expansion in the mixing terms:
%\begin{footnotesize}
\begin{equation}
\label{Mnu2}
\mathcal{M}_\nu=\left( \begin{array}{cc}
    A_2[v_{Li}][ v_{Lj}]/4
    + B_2[(h_\nu^{\prime \prime} )_{i}][v_{Lj}]/2 + C_2\left[ (h_\nu^{\prime \prime} )_{i} (h_\nu^{\prime \prime} )_{j}\right]
    &
    D_2  [v_{Li}] /2

\\
    D_2  [v_{Li}] /2  & E_2 \left [{1}/ {\mu} \right]

\end{array}
\right)
\end{equation}
%\end{footnotesize}
%
where, to first order in small mixing parameters, we have
\begin{eqnarray}
    A_2 & =& \frac{g_2^2}{M_{\tilde W}}-\frac{g_2M_{\tilde Y \tilde Y^\prime}}{v^2 \sin 2 \beta }~ , \\
    B_2 & =& -\frac{2} {v_u} \left ( 1+\frac {g_2^2M_{\tilde Y}}{g_Y^2M_{\tilde W}} \right )~,  \\
    C_2 &=& -\frac{4M_{\tilde Y}}{g_Y^2 v_u^2}~,\\
    D_2  &=& \left [  \left ( \frac{2\mu_u}{ v_u}+  \frac{g_2 \mu_d}{v_d} \right)  \left(  1-\frac {2g_2^2M_{\tilde Y}}{g_Y^2 M_{\tilde W}}\right )  \right] ~ ,\\
    E_2&=&- \left [ \left (\mu_d^2+\mu_u^2 \right ) \tan \beta+\mu_u^2 \frac{4M_{\tilde Y} \mu}{g_Y^2 v_u^2}+2\mu_u\mu_d \right ]~,
\end{eqnarray}
where we have further assumed $\mu_{u,d}^\prime \ll \mu, \mu_u, \mu_d$. The analytical results shown are approximate.  However, in the following section we shall perform exact numerical analyses for neutrino masses and mixing,  and include known solar and atmospheric mixing constraints.
\section{Numerical Analysis}
\label{sec:analysis}
%%%%%%%%%%%%%%%%%%%%%%%%%%%%%%%%%%%%%%%%%%%%%%%%%%

The neutrino mass matrix in both scenarios is diagonalized by a unitary matrix $U$
$$\mathbf{U_i}^T \mathcal{M}_\nu \mathbf{U_i}={\rm diag}(m_i).$$

The addition of one neutrino state leads to a generalization of the PMNS matrix to a $4\times4$ unitary matrix and introduces three new mass-squared differences $\Delta m_{4i}^2$ ($i=1,2,3$). Since $\Delta m_{41}^2 \gg \Delta m_{21,31}^2$, the [3+1] model effectively introduces four new parameters to neutrino oscillation phenomenology: one mass-squared difference $\Delta m_{41}^2$ and three angles $(\theta_{14},\theta_{24},\theta_{34})$ describing the active-sterile mixing. Assuming that all the CP-violating phases vanish, the $4\times4$ unitary mixing matrix $\mathbf{U_4}$ can be parametrized in the following way~\cite{deGouvea:2008nm}:
\begin{equation}\label{eq:para}
\mathbf{U_4}= \mathbf{R^{34}}(\theta_{34})\mathbf{R^{24}}(\theta_{24})\mathbf{R^{14}}(\theta_{14})\mathbf{R^{23}}(\theta_{23})\mathbf{R^{13}}(\theta_{13})\mathbf{R^{12}}(\theta_{12})~,
\end{equation}
where $\mathbf{R^{ij}}(\theta_{ij})$ ($i,j=1,\ldots,4$ and $i<j$) is the $4\times4$ rotation matrix in the $ij$-plane with the angle $\theta_{ij}$, with elements
\begin{equation}
\left[\mathbf{R^{ij}}(\theta_{ij})\right]_{kl}=(\delta_{ik}\delta_{il}+\delta_{jk}\delta_{jl})c_{ij}+(\delta_{ik}\delta_{jl}-\delta_{il}\delta_{jk})s_{ij}+\left[(1-\delta_{ik})(1-\delta_{jl})+(1-\delta_{il})(1-\delta_{jk})\right]\delta_{kl}~,
\end{equation}
where $c_{ij}\equiv\cos\theta_{ij}$ and $s_{ij}\equiv\sin\theta_{ij}$.

\begin{itemize}
\item
For normal hierarchy one assumes
$$m_{\nu_1}\le0.001~{\rm eV};~~m_{\nu_2}= \sqrt{\Delta m^2_{\rm sol} +m_{\nu_1}^2};~~m_{\nu_3}= \sqrt{\Delta m^2_{\rm atm} +m_{\nu_2}^2}~~{\rm and}~~m_{\nu_4}= \sqrt{\Delta m^2_{14} +m_{\nu_1}^2}.$$
\item For inverted hierarchy one assumes
$$m_{\nu_3} \le 0.001~{\rm eV};~~m_{\nu_2}= \sqrt{\Delta m^2_{\rm sol} +m_{\nu_1}^2};~~m_{\nu_1}= \sqrt{\Delta m^2_{\rm atm} +m_{\nu_3}^2}~~{\rm and}~~m_{\nu_4}= \sqrt{\Delta m^2_{34} +m_{\nu_3}^2},$$
\end{itemize}
with the constraints from solar and atmospheric neutrino mixings  \cite{Fogli:2012ua}
\begin{eqnarray}
7.27 \times 10^{-5} \le & \Delta m_{\rm sol}^2 & \le 8.03 \times 10^{-5} ~{\rm eV}^2, \\
2.17 \times 10^{-3} \le & |\Delta m_{\rm atm}^2| & \le 2.54 \times 10^{-3} ~{\rm eV}^2.
\end{eqnarray}
To obtain the correct neutrino masses and mixings, we scan for $U(1)^\prime$ parameters in the following ranges
\begin{eqnarray}
M_{\tilde Y}, M_{\tilde Y^\prime}, M_{\tilde W}, M_{\tilde Y \tilde Y^\prime} & \in & \pm  [0.5-5]~{\rm TeV}; \nonumber \\
\lambda, h_s, h_s^\prime  & \in & \pm [0- 1] ;  \nonumber \\
v_L\equiv \langle \tilde \nu _L\rangle& \in & [10^{-6}-10^{-4}]~{\rm GeV}; \nonumber \\
\tan \beta &\in &[1-30].
\end{eqnarray}
 For $M_R=10^{12}$ GeV,  $h_\nu \sim  10^{-12} \lambda$, though varying $\lambda$ is equivalent to varying $M_R$.
Additionally, in the first scenario, [3+1 RN], where we designate
the right-handed neutrino to be a candidate for light sterile
neutrino, and the singlino to be heavy, we have:
\begin{eqnarray}
v_S  \equiv \langle S \rangle& \in & [1-10^8] ~~{\rm TeV};  \nonumber \\
v_R \equiv \langle \tilde \nu_R\rangle& \in & [10^{-6}-10^{-2}]~{\rm GeV},
\end{eqnarray}
while in the second scenario [3+1 $\tilde S$], where we designate the singlino to be a candidate for light sterile neutrino and the right-handed neutrino to be heavy:
\begin{eqnarray}
v_S \equiv \langle S \rangle& \in &  [1-10^8] ~~{\rm TeV};  \nonumber \\
v_R \equiv \langle \tilde \nu_R\rangle& \in & [10^{3}-10^{6}]~{\rm GeV}.
\end{eqnarray}
The exact values of the masses and mixing parameters depend on the  $U(1)^\prime$
charges. For each model these are determined by the mechanism chosen to break from $E_6$ GUT symmetry to $SU(3)_c \times SU(2)_L \times U(1) \times U(1)^\prime$.   Gauge invariance of the superpotential require that the charges obey the conditions:
\begin{eqnarray}
Q^\prime_{H_d}+Q^\prime_{H_u}+Q^\prime_S &=& 0\\
Q^\prime_L+Q^\prime_{H_u}+Q^\prime_{N}&=&0.
 \end{eqnarray}
We assume, for simplicity, the $E_6$ GUT relation $g_{Y^\prime}= \sqrt{\frac53} g_Y$ but allow $M_{\tilde Y}$, $M_{\tilde Y \tilde Y^\prime}$ and  $M_{\tilde W}$ to vary independently.  We find that the numerical results are not sensitive to the specific values of these charges, allowing us to choose a $U(1)^\prime$ model with simple $Q^\prime$ assignments. In Table \ref{tab:Qprime} we give the values for the relevant charges $Q^\prime$ used in this model (corresponding to $x=-\frac29$ in Table \ref{tab:charge-sol}. The $U(1)^\prime$ charges given below set $g^\prime _1=g_{Y^\prime}= \sqrt {\frac53}\, g_2 \tan \theta_W$, which is the value used in our numerical scans.
\setlength{\voffset}{-0.5in}
\begin{table}[htbp]
 \begin{center}
\setlength{\extrarowheight}{-5.8pt} \small
\begin{tabular*}{0.95\textwidth}{@{\extracolsep{\fill}} cccccc}
\hline\hline
Field & $L$&$H_u$&$H_d$ &$N$ &$S$
 \\ \hline
 Charge $Q^\prime$& $1$ & $-2$ & $1$  & $1$ & $1$\\
 \hline\hline
\end{tabular*}
\caption{\sl\small The $U(1)^\prime$ charges used for the relevant particles in the model. } \label{tab:Qprime}
\end{center}
 \end{table}

In addition, we  allow the mixing angles between sterile and active neutrinos to be $10^{-5} \le \sin^2 \theta_{i4} \le 10^{-1}$,~(i=1, 2) \cite{Mirizzi:2013kva}, and restrict $m_{\nu_4} \approx 0.1-10$ eV \cite{Abazajian:2012ys}, the largest interval allowed by experiments.

\begin{figure}[htbp]
%\vskip -0.3in
\begin{center}$
    \begin{array}{ccc}
\hspace*{-1.5cm}
    \includegraphics[width=2.2in,height=2.5in]{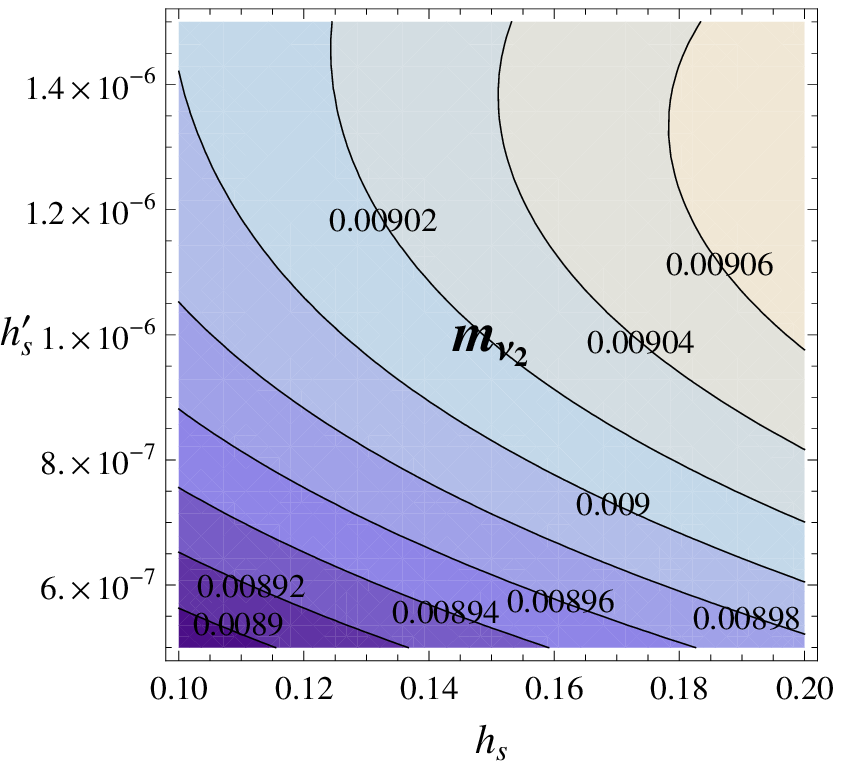}
&%\hspace*{-1.5cm}
  \includegraphics[width=2.2in,height=2.5in]{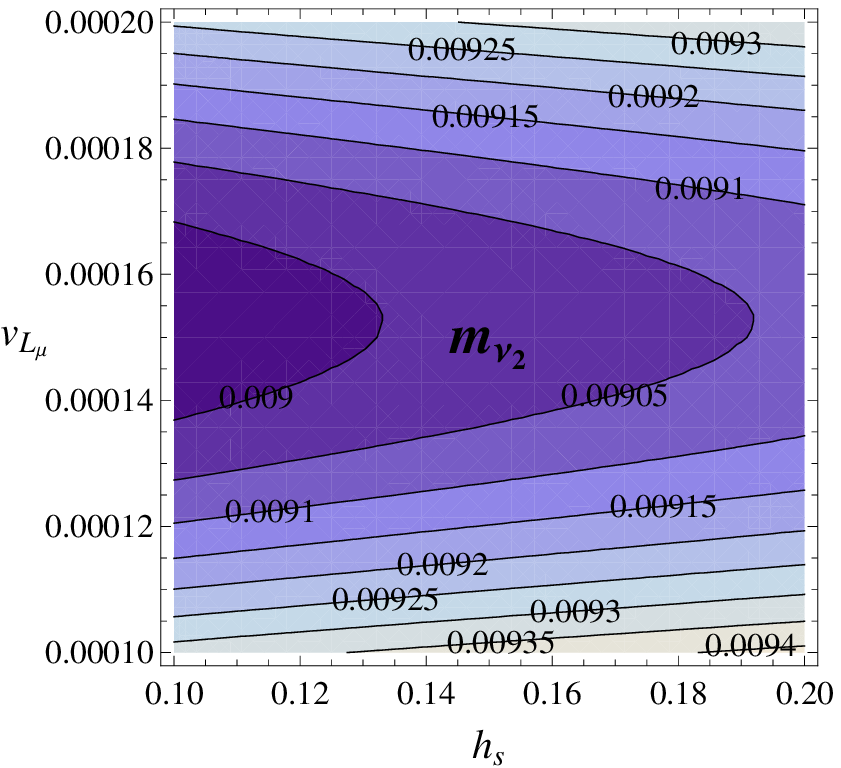}
    &\includegraphics[width=2.2in,height=2.5in]{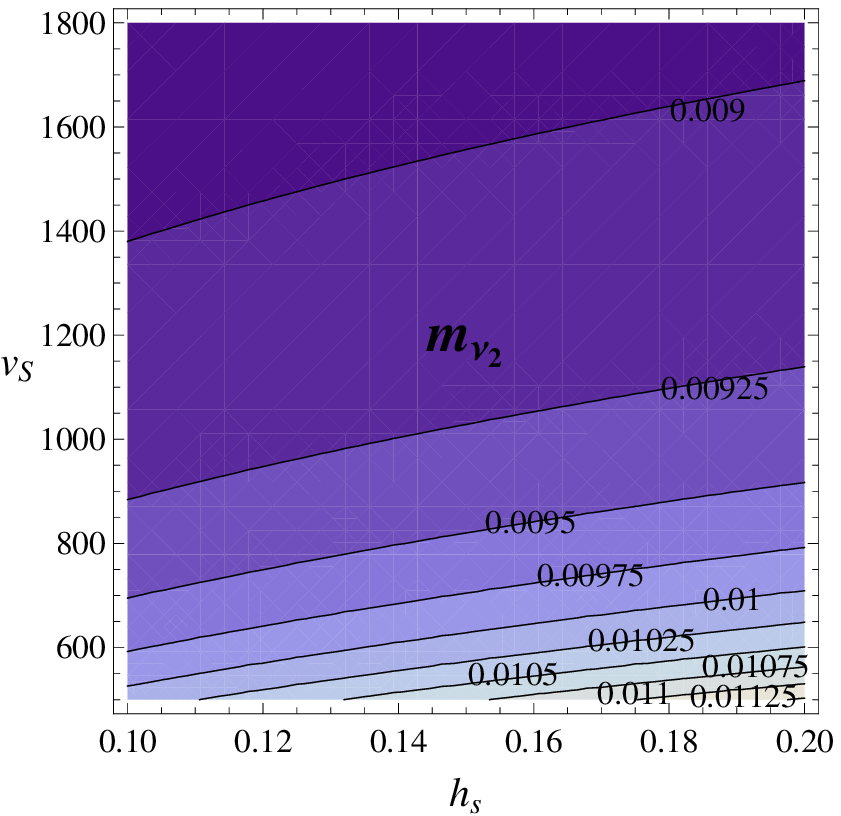}\\
    \hspace*{-1.5cm}
    \includegraphics[width=2.2in,height=2.5in]{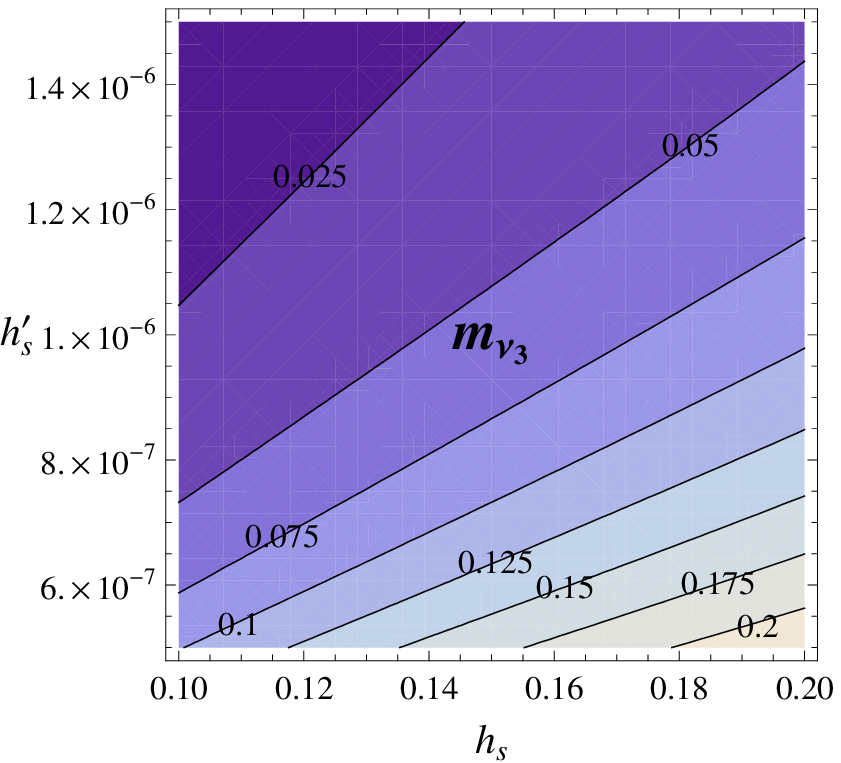}
  &  \includegraphics[width=2.2in,height=2.5in]{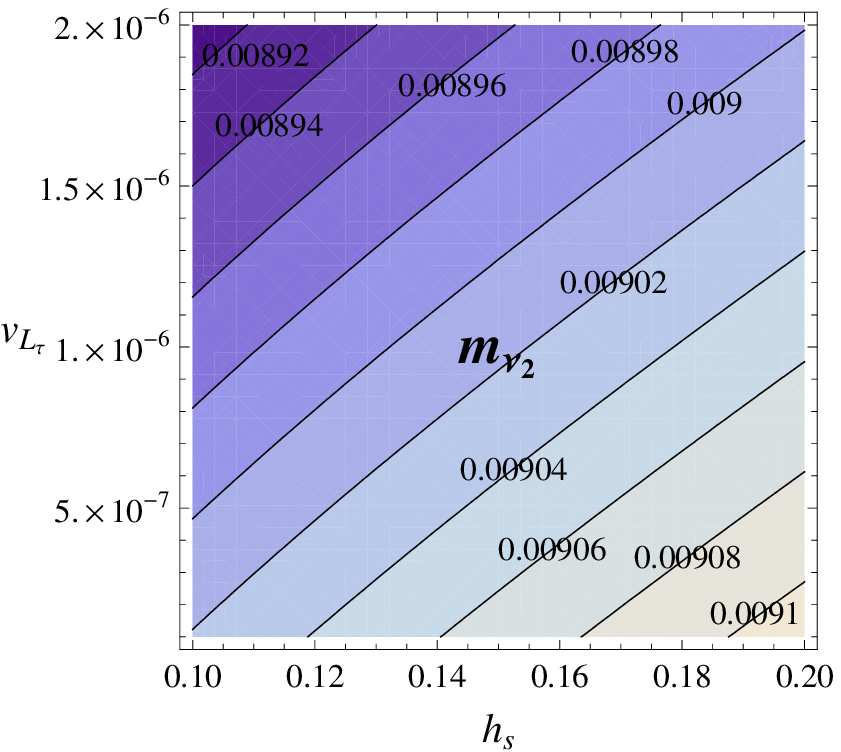}
&% \hspace*{-1.5cm}
    \includegraphics[width=2.2in,height=2.5in]{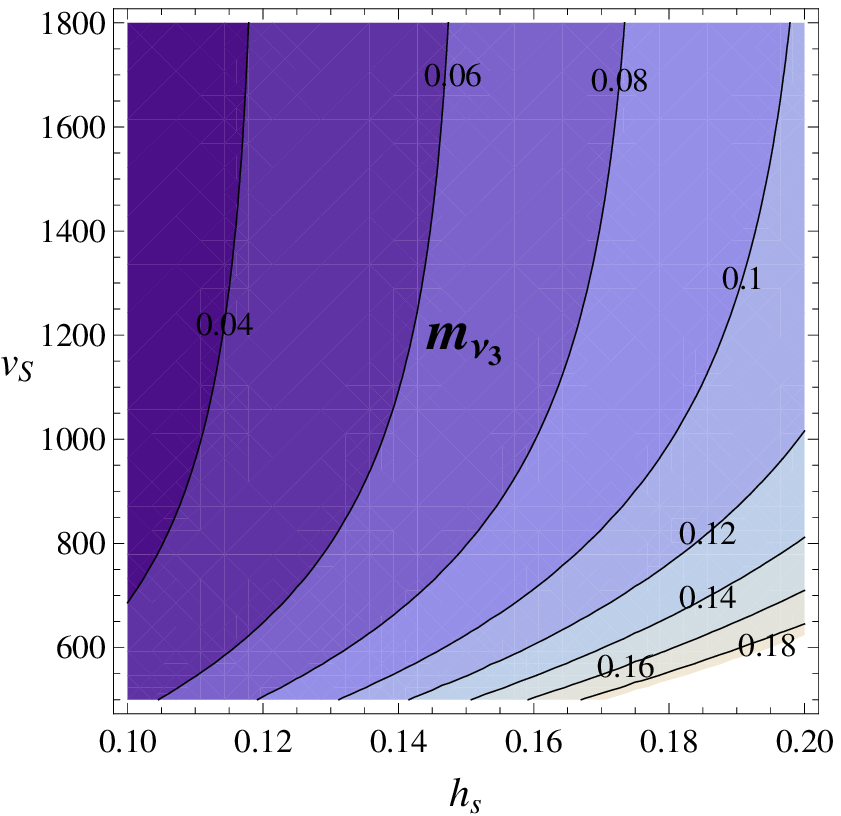}
     \end{array}$
\end{center}
\vskip -0.1in
 \caption{ (color online). \sl\small Contour plots for the variation of the mass of the $\nu_2$ and $\nu_3$ active neutrinos in the $h_s-h_s^\prime$  (left-hand column); the  contours of $m_{\nu_2}$ in the $h_s-v_{L_\mu}$  and in the $h_s-v_{L_\tau}$ plane (middle column); and contours of $m_{\nu_2}$ and $m_{\nu_3}$ the $h_s-v_S$ plane (right-hand column), for Scenario 1, [3+1 RN]. We fix the values for the solar and atmospheric neutrino mass splittings to their values within $1\sigma$.  Neutrino masses, as indicated on contours,  are in eV, while the sneutrino VEVs are in GeV.}
 \label{fig:3+1RN}
\end{figure}

We proceed by exploring the parameter space of the first scenario [3+1 RN]. We chose to plot graphs with either parameters that  yield significant  variations in the mass (such as $h_s,~h_s^\prime$), or highlight some typical parametric dependence, such as the contours in the $h_s-v_{L_\alpha}$ planes. We proceed as follows. We first scan over all the parameters to find allowed regions of the parameter space (14, to be exact: 3 $h_\nu$'s, 3 $v_L$'s, $h_s$, $h_s^\prime$, $v_R$, $v_S$, $M_{\tilde Y}, M_{\tilde W}, M_{\tilde Y^\prime}$ and $M_{\tilde Y \tilde Y^\prime}$) consistent with the neutrino data. We then vary the parameters in the allowed region to restrict parameters. We find that the neutrino masses and mixing are far more sensitive to some parameters than others. Thus we are confident that the allowed regions represent necessary restrictions on the parameter space and based on these, we are able to make some general comments regarding consequences of this scenario.  The plots shown represent the results of our global fits,  and  illustrate that within a simple scheme we can explain and fit  the data.  
In Fig. \ref{fig:3+1RN} we show contour plots for the variation of the mass of the $\nu_2$ and $\nu_3$ active neutrinos in the $h_s-h_s^\prime$  (left-hand column); the  contours of $m_{\nu_2}$ in the $h_s-v_{L_\mu}$ and in the $h_s-v_{L_\tau}$ planes (middle column); and contours of $m_{\nu_2}$ and $m_{\nu_3}$ the $h_s-v_S$ plane (right-hand column). Here $v_{L_\mu}$ is the VEV of the left-handed muon neutrino, and  $v_{L_\tau}$ is the VEV of the left-handed tau neutrino.  The masses vary with $v_S$, the VEV of the singlino, whereas they are almost independent of $v_R$, the VEV of the right-handed neutrino, in the allowed region.  In this scenario $v_R$ is light, as the right-handed neutrino is chosen to be the sterile candidate. 

An additional constraint on the spectrum is provided by the eigenvalues of the heavy neutralino mass $M$. The two lightest eigenstates of $M$ are degenerate mixtures of $\tilde H_u$ and $\tilde H_d$ higgsinos,  and  have masses $\sim $ 140 GeV, precluding decays of $Z$ boson into  higgsino pairs. These masses are sensitively dependent on the VEV $v_S$.
 For neutrinos, notice in particular the stringent constraints on the $h_s$($h_s^\prime$) parameters. These coupling parameters are responsible for generating  the effective $\mu$ parameter ($h_s$) and the effective $\mu^{\prime \prime}$ responsible for bilinear $R$-parity violation ($h_s^\prime$)  in Eq. (\ref{bilinear}).  We have chosen to show the variation of $m_{\nu_2}$ with the VEVs of the left-handed neutrinos; but similar plots exist for $m_{\nu_1}$ and $m_{\nu_3}$.
%%%%%%%%%%%%%%%%%%%%%%%%%%%%%%%%%%%%%%%%%%%%%%%%%%%%%%%%%%%%%%%%%%%%%%%%%%%%%%%%%%%%%%%%%%%%%%%%%%%%%
%%%%%%%%%%%%%%%%%%%%%%%%%%%%%%%%%%%%%%%%%%%%%%%%%%%%%%%%%%%%%%%%%%%%%%%%%%%%%%%%%%%%%%%%%%%%%%%%%%%%

\begin{figure}[htbp]
%\vskip -0.3in
\begin{center}$
    \begin{array}{ccc}
\hspace*{-1.5cm}
    \includegraphics[width=2.2in,height=2.5in]{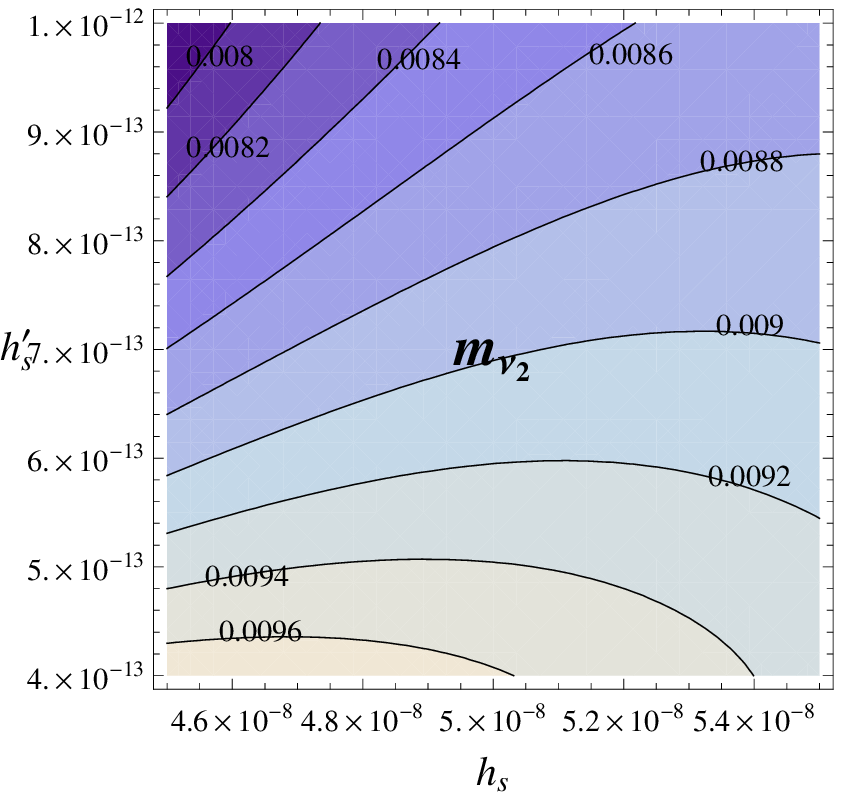}
&%\hspace*{-1.5cm}
  \includegraphics[width=2.2in,height=2.5in]{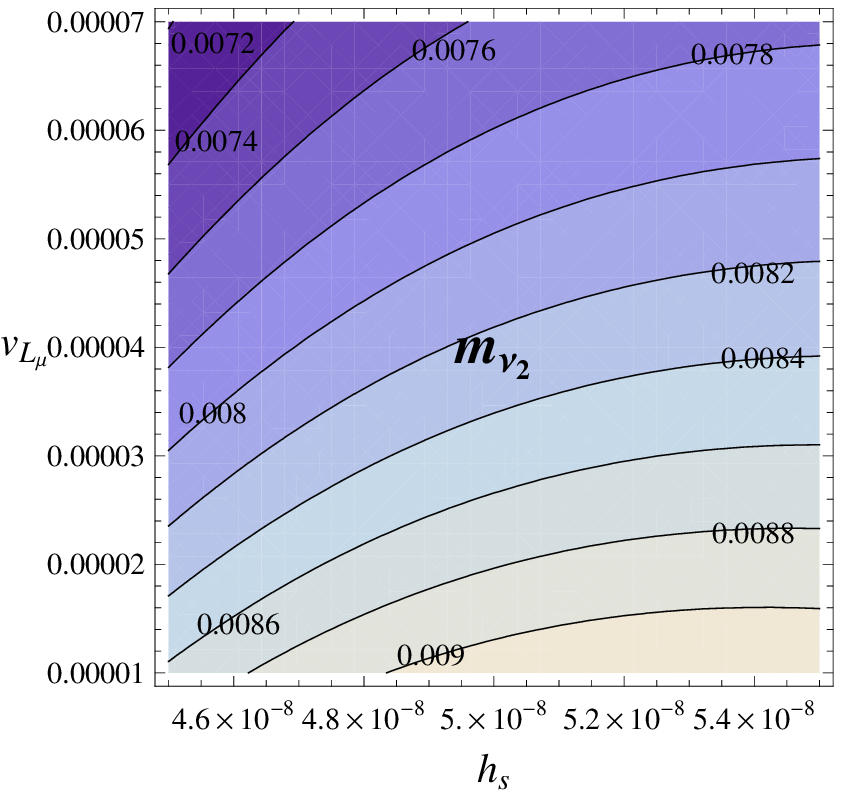}
    &\includegraphics[width=2.2in,height=2.5in]{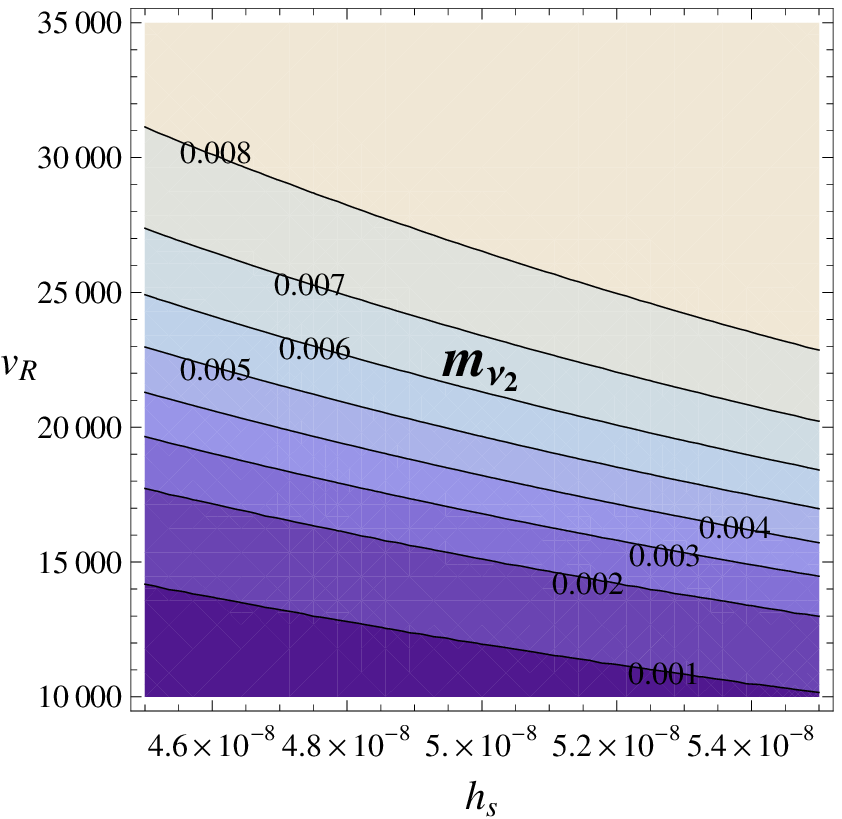}\\
    \hspace*{-1.5cm}
    \includegraphics[width=2.2in,height=2.5in]{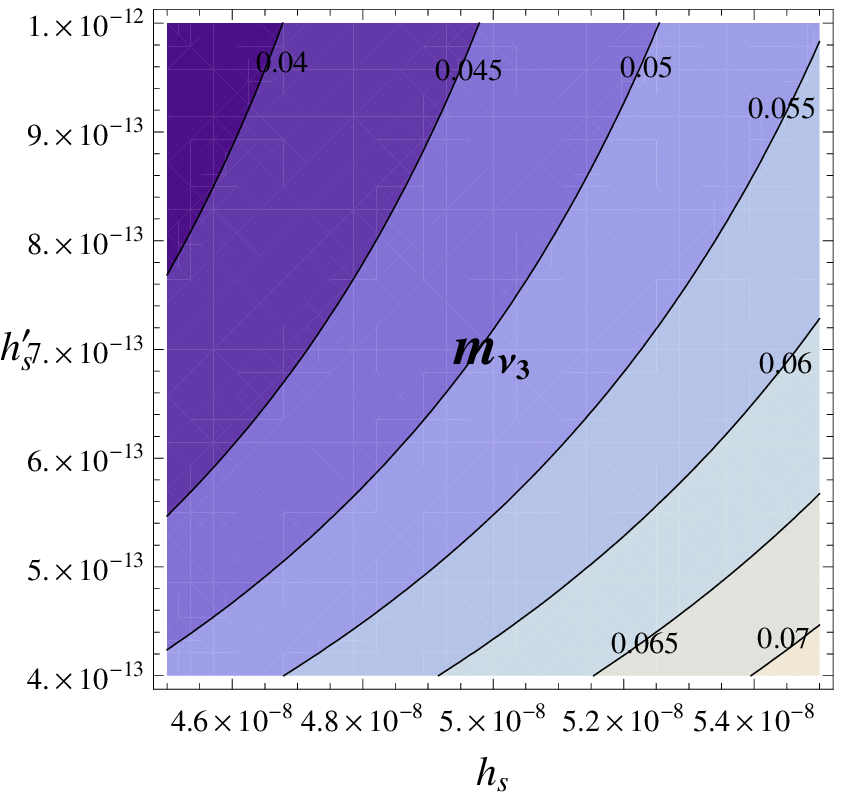}
  &  \includegraphics[width=2.2in,height=2.5in]{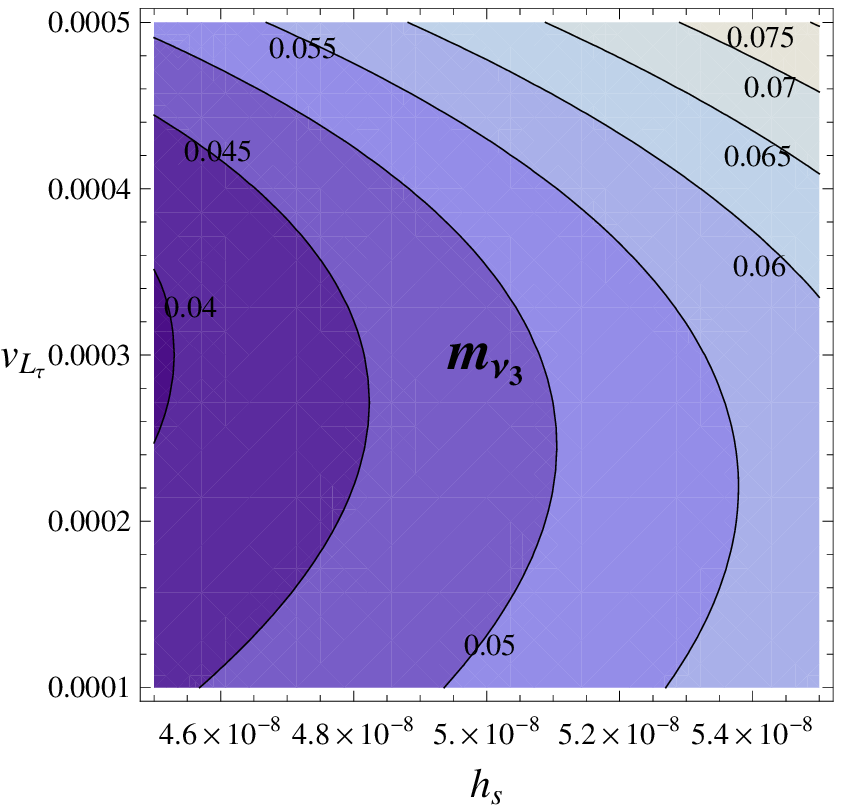}
&% \hspace*{-1.5cm}
    \includegraphics[width=2.2in,height=2.5in]{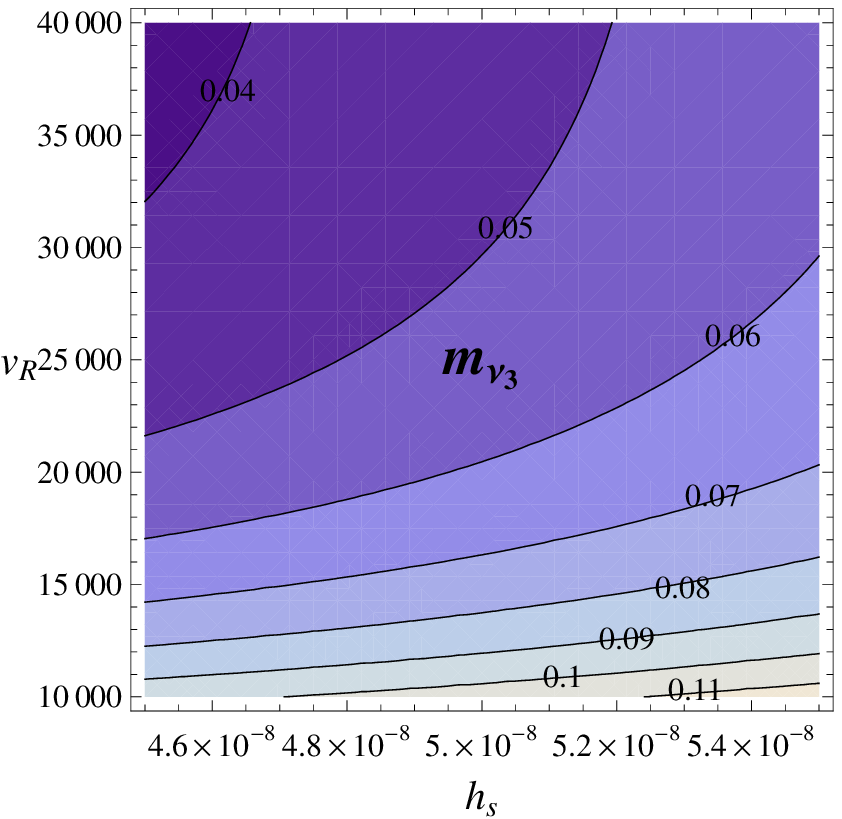}
     \end{array}$
\end{center}
\vskip -0.1in
 \caption{ (color online). \sl\small Contour plots for $m_{\nu_2}$ and $m_{\nu_3}$ in the $h_s-h_s^\prime$ plane (left-hand column);  of $m_{\nu_2}$ as a contour in the $h_s-v_{L_\mu}$ plane and of $m_{\nu_3}$ as a contour in the $h_s-v_{L_\tau}$ plane (middle column);  and of $m_{\nu_2}$ and $m_{\nu_3}$ in the $h_s-v_R$ plane, (right-hand column), for Scenario 2, [3+1 $\tilde S$] . We fix the values for the solar and atmospheric neutrino mass splittings to their values within $1\sigma$. Neutrino masses, as indicated on contours,  are in eV, while the sneutrino VEVs are in GeV.}
 \label{fig:3+1S}
\end{figure}

 We analyze next the same dependence on model parameters for the second scenario, where the sterile neutrino is the singlino, [3+1 $\tilde S$]. In Fig. \ref{fig:3+1S} we show the dependence of neutrino masses  $\nu_2$ and $\nu_3$   as contours in the $h_s-h_s^\prime$ plane (left-hand column);  of $m_{\nu_2}$ as a contour in the $h_s-v_{L_\mu}$ plane and of $m_{\nu_3}$ as a contour in the $h_s-v_{L_\tau}$ plane (middle column);  and of $m_{\nu_2}$ and $m_{\nu_3}$ in the $h_s-v_R$ plane (right-hand column). 
 
 As in the first scenario [3+1 RN], the plots are presented to illustrate that there exist regions of the parameter space which can fit the data.  We proceed the following way: we scan the parameter space till we find a region where constraints from neutrino masses and mixings are satisfied at 1$\sigma$, then vary the model parameters around this point.  We  identify  parameters which affect  the neutrino masses and mixings the most. The contours shown indicate  regions of the parameter space for which the constraints from the experimental data are satisfied.   Although many parameters are involved, small variations of only a few affect neutrino masses significantly. 
We can extract some  general features that emerge from the parameter scan and mass fit,  generic for all parameter points which satisfy both neutrino data and LHC constraints on neutralino masses.
 \begin{enumerate}
 \item The coupling $h_s$ which generates the $R$-parity conserving parameter $\mu$, and  the coupling $h_s^\prime$ which generates the bilinear R-violating parameter $\mu^{\prime \prime}$ are positive in both scenarios. 
  \item The  $v_R$ is a factor of $10^{10}$ larger in scenario [3+1 $\tilde S$] than in scenario [3+1 RN].  In the former $v_R$ is the VEV of the heavy right-handed neutrino, while in the latter it is the VEV of the light sterile neutrino. This is understood by the fact that, in the first scenario the right-handed neutrino is sterile and thus light, whereas in the second scenario it enables the seesaw mechanism.
  \item
 The values of the VEV $v_S$ required to satisfy the solar and atmospheric neutrino mass splittings are several orders of magnitude larger in [3+1 $\tilde S$] than those for scenario [3+1 RN], indicating that the $U(1)^\prime$ model is broken at a higher scale in that case ($v_S\sim 1$ TeV and $v_S \sim 10^6$ TeV for scenario 1 and 2, respectively). As the breaking scale in singlino scenario is unusually large for [3+1 $\tilde S$], we return to discussing it further after presenting the complete parameter sets.
   \item The $v_S$ values stem mostly from constraints on the mass eigenvalues for the heavy neutralino mass matrix, $M$. 
The bare masses of the $\tilde H_u$ and $\tilde H_d$ higgsinos are strictly proportional to $h_s v_S$. As in the [3+1 $\tilde S$] scenario the values of $h_s$ must be small to generate small neutrino masses, $v_S$ must be large to generate sufficiently large masses for the higgsinos, required to be heavier than $\sim 100$ GeV as per LEP constraints.  
 \item While in both scenarios $h_s^\prime$ is required to be small (as expected, as this is the $R$-parity violating coupling),  the parametric dependence of the masses is quite different and the ranges required for the parameters $h_s, h_s^\prime$ are a factor of $10^{-7}$ smaller in scenario [3+1 $\tilde S$] than  those in scenario  [3+1 RN].  This is the result of the fit (i.e., it is the only solution surviving all the constraints), and there is no {\it a priori} reason for this to be so. 
 \item The neutrino-generating Yukawa parameters also turn out to be a factor of $10^{-7}$ in scenario [3+1 ${\tilde S}$] than the corresponding values in [3+1 RN] scenario. They are required to be $h_\nu  \sim 10^{-12}$ in the first scenario, and $h_\nu  \sim 10^{-19}-10^{-20}$ in the second scenario. This {\it ad-hoc} hierarchical structure\footnote{This fine-tuned structure is typical of neutrino masses in split supersymmetry, see \cite{Chun:2004mu}.} has possible explanation that, on general grounds, we expect $h_\nu  \sim \lambda/M_R$ with $M_R$ a large mass scale, probably generated by the VEVs of the electron and muon sneutrinos, which are much larger than the mass scale of the heavy neutralinos, since they were assumed to decouple. It could be that in scenario [3+1 RN], the mass scale is expected to be $M_R\sim 10^{12}$ GeV, while in scenario [3+$\tilde S$] it is closer to the Planck scale $M_R \sim 10^{19}$ GeV, reflecting the difference of a factor of $10^6$ between  $U(1)^\prime$ breaking scales in the two scenarios. Alternatively, the mass scale $M_R$ could be the same in both scenarios, while the coupling $\lambda$ could be fine-tuned to be a factor of $10^{-7}$ smaller in scenario 2, in agreement with the fine-tuning of all Yukawa couplings.
\item
 In both scenarios the two lightest  eigenvalues of $M$ are even mixtures of $\tilde H_u$ and $\tilde H_d$ and have masses of 140 GeV in scenario 1, and 117 and 120 GeV, respectively in scenario 2\footnote{Note that the neutralinos can have masses below $M_Z/2$ as long as they are bino-like.}. Note that in this case the neutral higgsinos are close in mass to the charged higgsinos and thus we require them to be heavier than $\sim 100$ GeV, the center-of-mass energy at LEP. Of course, these are no longer stable LSPs and are expected to decay, and, as they couple strongly to the $Z$ boson, they might be seen the LHC.
\item Neutrino masses are very weakly dependent on the sterile neutrino VEV (in the first case $v_R$, in the second case $v_S$), for  sterile neutrino VEVs in the allowed region by solar and atmospheric mass constraints. 
\item In both scenarios, the sterile-active neutrino mass-squared splitting is of order 2-5 eV$^2$, and cannot be decreased further.
\end{enumerate}
%%%%%%%%%%%%%%%%%%%%%%%%%%%%%%%%%%%%%%%%%%%%%%%%%%%%%%%%%%%%%%%%%%%%%%%%%%%%%%%%%%%%%%%%%%%%%%%%%%%%%
%%%%%%%%%%%%%%%%%%%%%%%%%%%%%%%%%%%%%%%%%%%%%%%%%%%%%%%%%%%%%%%%%%%%%%%%%%%%%%%%%%%%%%%%%%%%%%%%%%%%%%
%
While the figures show regions of allowed parameter space, in Table \ref{tab: neutrinoparam}
 we list the masses and mixing parameter values for a point representative of our fit for scenarios [3+1 RN] and [3+1 $\tilde S$], and, by comparison, the results of the fit from the data from MiniBooNE, MiniBooNE and Gallium, and PLANCK.  The PLANCK column includes previous bounds on neutrino masses and mixings, thus showing after-PLANCK parameter restrictions. We note that, while future data would be able to tighten the restrictions on the parameters further, the dependence on parameters in both scenarios is very tight, and, for instance, we were unable to fit the data in either scenario with smaller $\Delta m^2_{41} $ mass splitting. In fact, despite significant differences in the parameter space, there are no substantial differences between neutrino masses and mixings in the two scenarios. The mixing angles,  indicative of neutrino oscillations, in Table  \ref{tab: neutrinoparam} are defined as
 \begin{eqnarray}
 \sin^2 2 \theta_{\alpha \beta}&=&4 |U_{\alpha4}|^2 |U_{\beta 4}|^2~, \nonumber \\
 \sin^2 2 \theta_{\alpha \alpha}&=&4 |U_{\alpha4}|^2\left ( 1- |U_{\alpha 4}|^2 \right)~,
\end{eqnarray}
for $\alpha, \beta=e, \mu, \tau, s$.  
%%%%%%%%%%%%%%%%%%%%%%%%%%%%%%%%%%%%%%%%%%%%%%%%%%%%%%%%%%%%%%%%%%%%%%%%%%%%%%%%%%%%%%%%%%%%%%%%%%%%%
\setlength{\voffset}{-0.5in}
\begin{table}[htbp]
 \begin{center}
\setlength{\extrarowheight}{-5.8pt} \small
\begin{tabular*}{0.95\textwidth}{@{\extracolsep{\fill}} cccccc}
\hline\hline
$\rm Parameter$ & $ \rm {MiniBooNE}$  \cite{Giunti:2011hn}&$\rm {MiniBooNE+GAL}$ \cite{Giunti:2011hn}  &$\rm {PLANCK}$  \cite{Mirizzi:2013kva} & $\rm {3+1~RN} $&$\rm {3+1~\tilde S}$\\
 \cline{1-6}\cline{1-6}
  $\Delta m^2_{21}/10^{-5} $[eV$^2$] &$7.54$&$7.54$&$7.5$&$8.0$&$7.7$\\
   $\Delta m^2_{31}/10^{-3} $[eV$^2$] &$2.4$&$2.42$&$2.43$&$2.3$&$2.3$\\
 $\Delta m^2_{41} $[eV$^2$] &$5.6$&$5.6$&$>1$&$3.6$& $2.6$\\
$|U_{e4}|^2  $&$0.032$&$0.037$&$10^{-5}-10^{-1}$&$1.8 \times 10^{-4}$&$ 6 \times 10^{-3}$\\
$|U_{\mu 4}|^2$ &$0.014$&$0.012$&$10^{-5}-10^{-1}$&$7.7 \times 10^{-3}$&$6 \times 10^{-4}$\\
$\sin^2 2\theta_{e \mu}$&$0.0018$&$0.0018$&$<10^{-2.5}$&$5.6 \times 10^{-6}$&$1 \times 10^{-5}$\\
$\sin^2 2\theta_{ee}$&$0.12$&$0.14$&$<10^{-2.5}$&$7.2 \times 10^{-4}$&$2.3 \times 10^{-2}$\\
$\sin^2 2\theta_{\mu \mu }$&$0.054$&$0.049$&$<10^{-2.5}$&$0.03$&$0.002$\\
\hline \hline
\end{tabular*}
\caption{\sl\small Parameter values for neutrino oscillations in the 3+1 fit, from data at MiniBooNE, MiniBooNE and Gallium, PLANCK, and the two sterile neutrinos presented in this work: the right handed neutrino as sterile neutrino, 3+1 RN; and the singlino as sterile neutrino, 3+1 $\tilde S$.} \label{tab: neutrinoparam}
\end{center}
 \end{table}

 Finally, in Table \ref{tab: cp inputs} we present the complete list of the values for model parameters corresponding to the fit from Table \ref{tab: neutrinoparam}. This set is characteristic of the parameter points shown in Figures \ref{fig:3+1RN}-\ref{fig:3+1S} which satisfy our constraints. The low value of $\tan \beta \approx 1$ is enforced by  constraints from $B_s \to \mu^+ \mu^-$ branching ratio \cite{Aaij:2012nna}, as well as by requiring a light CP-even SM-like Higgs boson with (tree-level) mass $m_{H^0}\approx 125$ GeV \cite{Frank:2013yta}.  Notice that the main difference between the two scenarios lies in the allowed values for $v_R$ ($\sim 10^{-6}$ GeV for the RH neutrino as sterile neutrino, $10^4$ GeV for the case of singlino as sterile neutrinos), $v_S$ (1.5 TeV for [3+1 RN], ${\cal O}(10^6$ TeV) for the [3+1 $\tilde S$]) and the couplings $h_s,~ h_s^\prime$ and $h_\nu$, as discussed before.

 In the  [3+1 $\tilde S$] scenario, physics is intrinsically fine-tuned,  since it requires a large $v_S$ but small Yukawa couplings to satisfy the sterile neutrino constraints, the symmetry breaking, and generation of $\mu_{eff}$. The requirement of such a large $v_S$ is similar to the case of split supersymmetry\footnote{Note that even the value of $v_S$ is similar, $10^9$ GeV in the singlino scenario, $10^{9}$ GeV in split supersymmetry with gauge coupling unification.} \cite{ArkaniHamed:2004fb}. Split supersymmetry eliminates the requirement of naturalness to avoid fine-tuning, and pushes the scale of supersymmetry breaking to $M_{\rm SUSY} \gg 1$ TeV. The advantage is that it evades many of the phenomenological constraints which affect generic supersymmetric extensions of the SM. In scenario [3+$\tilde S$], as in split supersymmetry, by widening the supersymmetry-breaking scale between the scalar and the gaugino sector, the squarks and sleptons become heavy, while charginos and neutralinos remain still  at the TeV scale or below.  As in split supersymmetry, in scenario [3+ ${\tilde S}] $ the low-energy effective theory is particularly simple. In addition to the Standard Model spectrum including the Higgs boson, the only extra particles are the neutralinos,  charginos, and a gluino, which is is long-lived. The Higgs doublet masses are light, and so are the associated higgsinos, which are accessible the LHC and the ILC. Split supersymmetry with  $R$-parity violation (including both  bilinear and the trilinear terms)  has been used before to explain neutrino masses \cite{Chun:2004mu}, with solar neutrino masses generated at one-loop (usually involving heavy Higgs bosons), and atmospheric ones at tree level. The requirement that $v_S \sim 10^9$ GeV from scenario [3+ ${\tilde S}]$ is completely consistent with the findings in \cite{Chun:2004mu}, and so is the condition for the smallness of the  bilinear parameters. Of course, our model has the additional advantage of solving the $\mu$ problem and explaining proton stability, as well as introducing a (new) candidate for the sterile neutrino.

 There are other differences in orders of magnitude and signs among $M_{\tilde W},~M_{\tilde Y},~M_{\tilde Y^\prime}$ and $M_{\tilde Y \tilde Y^\prime}$, but these characterize more the precise details of the fit than general constraints on the masses.
%{\bf TABLE for [3+1] Scenario}
%%%%%%%%%%%%%%%%%%%%%%%%%%%%%%%%%%%%%%%%%%%%%%%%%%%%%%%%%%%%%%%%%%%%%%%%%%%%%%%%%%%%%%%%%%
\setlength{\voffset}{-0.5in}
\begin{table}[htbp]
 \begin{center}
\setlength{\extrarowheight}{-5.8pt} \small
\begin{tabular*}{0.95\textwidth}{@{\extracolsep{\fill}} ccc}
\hline\hline
 $\rm Parameters$ & $\rm {3+1~RN} $  &$\rm {3+1~\tilde S}$
 \\ \cline{1-2}\cline{1-3}
 $\tan \beta$&$1.01$&$1.01$\\
   $h_{s}$&$0.130$&$5\times10^{-8}$\\
    $h_{s}^{\prime}$&$9.5\times10^{-7}$&$8\times10^{-13}$\\

   $h_{\nu_{e}}$&$10^{-12}$&$7.5 \times10^{-19}$\\
   $h_{\nu_{\mu}}$&$10^{-12}$&$9 \times10^{-19}$\\
   $h_{\nu_{\tau}}$&$10^{-12}$&$10^{-20}$\\

   $v_{L_{e}}$ [GeV]&$3\times10^{-4}$&$2\times10^{-5}$\\
   $v_{L_{\mu}}$ [GeV]&$1.5\times10^{-4}$&$2\times10^{-5}$\\
   $v_{L_{\tau}}$ [GeV]&$1\times10^{-6}$&$2.9 \times10^{-4}$\\
   $v_R$ [GeV]&$1.2 \times10^{-6}$&$3.3 \times10^{4}$\\
   $v_S$ [GeV]&$1500$&$3.4 \times10^{9}$\\

   $M_{\tilde W}$ [GeV]&$-1200$&$~1720$\\
   $M_{\tilde Y}$ [GeV]&$-300 $&$-1307 $\\
   $M_{\tilde Y \tilde Y^\prime}$ [GeV]&$~500$&$~1000$\\
   $M_{\tilde Y^\prime}$ [GeV]&$-1500$&$-2614$\\
   \hline\hline
   \end{tabular*}
\caption{\label{tab: cp inputs}\sl\small The parameter values for Scenario 1 [3+1 RN] and Scenario 2 [3+1 $\tilde S$] corresponding to the fit in Table \ref{tab: neutrinoparam}.}
\end{center}
 \end{table}
%%%%%%%%%%%%%%%%%%%%%%%%%%%%%%%%%%%%%%%%%%%%%%%%%%%%%%%%%%%%%%%%%%%%%%%%%%%%%%%
 %%%%%%%%%%%%%%%%%%%%%%%%%%%%%%%%%%%%%%%%%%%%%%%%%%%%%%%%%%%%%%%%%%%%%%%%%%%%%%%%%%%%%%%%%%%%%%%%%%%%%%

 \newpage

 \section{Summary and Conclusion}
 \label{sec:conclusion}

 The existence of sterile neutrinos  is an interesting development in the study of physics beyond the SM. Sterile or inert neutrinos have been suggested to provide a solution to some, but by no means all, neutrino experiments as a means of explaining apparently anomalous oscillations, and they appear to fit some of the data. Since sterile neutrinos are also favored by some of the cosmological data, their role as a harbinger for new physics has been explored extensively in the literature, even though the theoretical foundation for their existence is not very well motivated.
 In this work, we introduce a novel candidate for sterile neutrinos in the $[3+1]$ mixing scenario (three active, one sterile), and compare it with a popular scenario, in the context of $R$-parity breaking $U(1)^\prime$ extended supersymmetric models.  The model, as implemented in our work, presents a resolution to both the $\mu$-problem and the proton decay, both serious issues in MSSM. Thus, even without sterile neutrino candidates, this model provides a solid framework for phenomenology. We consider neutrino mixing schemes in which the three standard neutrinos have masses smaller than 1 eV and the (one) additional sterile neutrino has mass at the eV scale.  Notwithstanding, other sterile neutrinos may exist. While the existence of more sterile neutrinos which have been thermalized in the early universe comes into conflict with Big-Bang Nucleosynthesis data and measurements from the Cosmic Microwave Background, these cosmological constraints can be avoided by suppressing the thermalization of sterile neutrinos in the early universe. One can then consider an additional sterile neutrino with mass at the keV scale as a candidate for Warm Dark Matter, but stability and production mechanisms constraints must be satisfied. This scenario is beyond the scope of the present analysis.

 When $R$-parity is broken and lepton violation is allowed, baryon number violating terms are forbidden. We choose a framework in which lepton number is violated  through bilinear terms, or spontaneously, by allowing sneutrinos to acquire a VEV. The neutrino and neutralino sector mix, and two natural candidates for sterile neutrinos emerge: either the right-handed neutrino, or the singlino, both singlets with no couplings to the $Z$-boson. 
 
 For the case in which the right handed  neutrino plays the role of sterile neutrino, this neutrino must be light  (we chose only one, the $\tau$ right-handed neutrino for simplicity). In this scenario, the light right-handed neutrino mixes with the (active) left-handed neutrinos,  the singlino is heavy, and the rest of the neutralino mass matrix serves as a seesaw mechanism to yield small masses and mixings in the neutrino sector. This scenario resembles MSSM with $R$-parity violation, with the addition of a  sterile neutrino. The $U(1)^\prime$ breaking scale is $\sim 1.5$ TeV, which is within the limits of LHC.  The sterile neutrino has mass  $m_4 \sim 2$ eV. There are four neutralinos with masses of ${\cal O}(10^2$ GeV), the lightest of which are two degenerate, even mixtures of doublet higgsinos, $\tilde H_u$ and $\tilde H_d$. These should be produced at sufficient rates at the LHC, however their decay spectrum would likely contain only soft energy and thus these could be much easier to observe at the ILC with $\sqrt{s}\ge 300$ GeV. 
 %\cite{Baer:2013vqa}. 

In the second scenario in the $R$-parity breaking $U(1)^\prime$ model, the singlino  can act as a novel candidate for the sterile neutrino. When this occurs, the right-handed neutrino is heavy and becomes part of the heavy neutralino mixing matrix, effectively reinforcing the same seesaw mechanism as before.  This model is very different from the model in scenario [3+1 RN]. The $U(1)^\prime$ symmetry is broken at a very high scale, $v_S \sim 10^6$ TeV and this scenario is effectively a split supersymmetry scenario. The implications for the mass of the vector boson $Z^\prime$ and the singlet Higgs $S$ are that these will be beyond the LHC scale, not inconsistent with the present LHC data. Thus in the non-supersymmetric part, the model will resemble SM. In the SUSY sector, the squarks and sleptons are heavy, the fermionic partner of the singlet Higgs is very light $m_4 \sim 1.6$ eV. The only ${\cal O}(10^2$ GeV) neutralinos are Higgsinos, with masses very close to  the LEP limit,  $m_{\tilde H} \sim 117, 120$ GeV.   In order to satisfy both the neutrino mass data and the neutralino mass limits, this scenario requires very small neutrino-generating couplings, $h_\nu \sim \lambda/M_R \sim 10^{-19}$, reflecting either coupling fine-tuning inherent in split supersymmetry, or a large scale $M_R$.  

In both cases, we impose conditions of mass splittings compatible with solar and atmospheric neutrino experiments at $1\sigma$. We then compare mass splittings and mixings with recent experiments: MiniBooNE, MiniBooNE and Gallium, and PLANCK. The results of the two scenarios are similar in their prediction of $\Delta m_{41}^2$ mass splitting (albeit, for very different values of the parameters in the model). That is, $\Delta m_{41}^2 \sim 2-4$ eV$^2$, and should further neutrino experiments definitely require $\Delta m_{41}^2<10^{-1}$ eV$^2$, as cosmological hints seem to suggest, neither scenario would survive. The other definite prediction of both scenarios is the existence of two light higgsinos, with masses near the LHC limits, which could be visible at the LHC, and more likely at the ILC. This is a generic prediction of the model, resulting from requiring to satisfy both constraints from neutrino experiments and collider bounds.

Distinguishing between the two scenarios presented in this work could come from production and decays of the heavy neutralinos. In both scenarios, the lightest neutralinos are even mixtures of $\tilde H_u$ and $\tilde H_d$, which couple to the $Z$ boson, and  may be seen at the LHC.  In the first scenario, where the right-handed $\tau$ neutrino is sterile, the bino has mass $\sim 230$ GeV, and  the (mostly) singlino has mass $\sim 350$ GeV and  would show a distinctive decay pattern, as unlike the bino the singlino does not couple directly to fermions; the rest of the neutralinos are in the TeV region. The higgsinos would be copiously produced in cascade decays of squarks and gluinos.

In the second scenario where the singlino is the sterile neutrino, there are only two remaining neutralinos below the TeV scale: the two doublet higgsinos $\tilde H_u$ and $\tilde H_d$. This again is different from  the spectrum of MSSM where the bino is the lightest neutralino. In this scenario, the higgsinos will be produced through Drell-Yan, as in split supersymmetric models, and not through squark or gluino production. They will decay further, as $R$-parity is violated, hopefully to a long lived lightest supersymmetric particle. The measurement of the Higgs-higgsino-gaugino coupling, most likely at the ILC, is considered a promising test of high scale supersymmetry.

In conclusion, while the existence of sterile neutrinos is by no means a {\it fait-accompli}, and does not explain all neutrino data, it remains an exciting possibility for physics beyond the SM. The model and scenarios presented here are highly constrained, and future data  from planned experiments (reactor, short-baseline, source, decay-at-rest) would soon be able to confirm or rule them out.
%%%%%%%%%%%%%%%%%%%%%%%%%%%%%%%%%%%%%%%%%%%%%%%%%%%%%%%%%%%%%%%%%%%%%%%%%%%%%%%%%%%%%%%%%%%%%%%%%%%%%%
\section{Acknowledgments}
 The
work of  M.F.  is supported in part by NSERC under grant number
SAP105354. The research of L. S. is  supported in part by The
Council of Higher Education of Turkey (YOK).

\end{document}